\documentclass[prd,twocolumn,showpacs,preprintnumbers,amsmath,amssymb,superscriptaddress,amsmath]{revtex4-1}

\makeatletter
\renewcommand*\env@matrix[1][c]{\hskip -\arraycolsep
  \let\@ifnextchar\new@ifnextchar
  \array{*\c@MaxMatrixCols #1}}
\makeatother

\usepackage{amsmath,array}
\usepackage{graphicx,subfigure}  % Include figure files
\usepackage{dcolumn}                 % Align table columns on decimal point
\usepackage{bm}                          % bold math

\def\nuebar{\bar{\nu}_e}
\graphicspath{{./figures/}}

\begin{document}

%\preprint{Physical Review D}
  
\title{The Reactor Antineutrino Anomaly}

\author{G.~Mention}
\affiliation{CEA, Irfu, SPP, Centre de Saclay, F-91191 Gif-sur-Yvette, France}

\author{M.~Fechner}
\affiliation{CEA, Irfu, SPP, Centre de Saclay, F-91191 Gif-sur-Yvette, France}

\author{Th.~Lasserre}
\email{Corresponding author: thierry.lasserre@cea.fr}
\affiliation{CEA, Irfu, SPP, Centre de Saclay, F-91191 Gif-sur-Yvette, France}
\affiliation{Astroparticule et Cosmologie APC, 10 rue Alice Domon et L\'eonie Duquet, 75205 Paris cedex 13, France}

\author{Th.~A.~Mueller}
\affiliation{CEA, Irfu, SPhN, Centre de Saclay, F-91191 Gif-sur-Yvette, France}

\author{D.~Lhuillier}
\affiliation{CEA, Irfu, SPhN, Centre de Saclay, F-91191 Gif-sur-Yvette, France}

\author{M.~Cribier}
\affiliation{CEA, Irfu, SPP, Centre de Saclay, F-91191 Gif-sur-Yvette, France}
\affiliation{Astroparticule et Cosmologie APC, 10 rue Alice Domon et L\'eonie Duquet, 75205 Paris cedex 13, France}

\author{A.~Letourneau}
\affiliation{CEA, Irfu, SPhN, Centre de Saclay, F-91191 Gif-sur-Yvette, France}

\date{\today}

\begin{abstract}
Recently, new reactor antineutrino spectra
have been provided for $^{235}$U, $^{239}$Pu, $^{241}$Pu, and $^{238}$U, 
increasing the mean flux by about 3~percent.
To a good approximation, this reevaluation applies to
all reactor neutrino experiments. 
The synthesis of published experiments at reactor-detector distances
$<100$~m leads to a ratio of observed event rate to predicted rate of
$0.976\pm 0.024$.
With our new flux evaluation, this ratio shifts to $0.943\pm 0.023$,
leading to a deviation from unity at 98.6\%~C.L. which we call 
the reactor antineutrino anomaly.
The compatibility of our results with the existence of a fourth
non-standard neutrino state driving neutrino oscillations at short
distances is discussed. The combined analysis of
reactor data, gallium solar neutrino calibration experiments, 
and MiniBooNE-$\nu$ data disfavors the no-oscillation hypothesis
at 99.8\%~C.L. The oscillation parameters are such that
$|\Delta m_{\rm new}^2 | > 1.5$~eV$^2$~(95\%) and 
$\sin^2(2\theta_{\rm new})=0.14\pm0.08$~(95\%). 
Constraints on the $\theta_{13}$ neutrino mixing angle are revised.
\end{abstract}

\maketitle

\section{Introduction}
Neutrino oscillation experiments over the last twenty
years have established a picture of neutrino mixing
and masses that explains the results of solar, atmospheric
and reactor neutrino experiments~\cite{pdg}.
These experiments are consistent with the mixing of
$\nu_{e}$, $\nu_{\mu}$ and $\nu_{\tau}$ with three mass eigenstates,
$\nu_1$, $\nu_2$ and $\nu_3$.  In particular, the mass differences
are required to be
$|\Delta m_{31}^2|\simeq2.4\,10^{-3}~{\rm eV}^2$ and
$\Delta m_{21}^2/|\Delta m_{31}^2|\simeq 0.032$.

Reactor experiments have played an important role in
the establishment of this pattern. In experiments at distances $<100$~m
from the reactor core, at ILL-Grenoble, Goesgen, Rovno, Krasnoyarsk, Savannah River
and Bugey~\cite{ILL81, Goesgen,Rovno88, Rovno91,Krasnoyarsk_87,Krasnoyarsk_94,Bugey3,Bugey4,SRP96}, 
the measured rate of $\nuebar$ was found to be in reasonable agreement with 
that predicted from the reactor antineutrino spectra, 
though slightly lower than expected, with the measured/expected 
ratio at $0.976\pm 0.024$ (including recent revisions of the neutron mean lifetime). 
This opened the way to longer baseline experiments measuring the 
$\nuebar$ survival probability $P_{ee}$ at various distances.  

Including the three known active neutrinos, $P_{ee}$ can be written
in first approximation as
\begin{equation}
P^{2 \nu}_{ee}(L)=1-\sin^2(2\theta_{i})\sin^2\left(1.27\frac{\Delta
m_{i}^2[{\rm eV}^2]L[{\rm m}]}{E_{\nuebar}[{\rm MeV}]}\right), 
\end{equation}
where $\Delta m_{i}^2=\Delta m_{21}^2$ and~$\theta_{i}\sim\theta_{12}$
if the baseline exceeds a few tens of kilometers, and $\Delta m_{i}^2=\Delta m_{31}^2$
and $\theta_{i}=\theta_{13}$ if it does not exceed a few
kilometers~\cite{cras2005}. 

The CHOOZ experiment was located 1~km away from two 4.3~GW 
reactors~\cite{Chooz}, and did not report any neutrino oscillation in
the parameter region~$\Delta m_{31}^2>10^{-3}~{\rm eV}^2$.
In addition CHOOZ data provides the world's best constraint on 
the~$\theta_{13}$ mixing angle value, at $\sin^2(2\theta_{13})<0.14$ 
(90\%~C.L.) for $|\Delta m_{31}^2|\simeq 2.4\,10^{-3}~{\rm eV}^2$. 
Forthcoming kilometer baseline experiments with multiple detectors are 
underway to provide a clean measurement of the last undetermined
neutrino mixing angle~$\theta_{13}$~\cite{Dayabay,Doublechooz,Reno}.  

The KamLAND experiment detected electron 
antineutrinos from surrounding reactors located at an average
distance of 180 km, thus probing the $\Delta
m_{21}^2\sim 10^{-4}~{\rm eV}^2$ region. KamLAND successfully reported a
strong oscillation signal in agreement with solar neutrino data~\cite{Kamland}.
A three-flavor oscillation treatment shows that KamLAND also has
a weak sensitivity to the~$\theta_{13}$ mixing angle, as 
$P_{ee} \sim \cos^4\theta_{13} P^{2 \nu}_{ee}$, leading to the
anti-correlation of $\sin^2\theta_{13}$ and $\sin^2\theta_{12}$.
Recent global fits including solar neutrino and reactor oscillation data
indicated a preference for non-zero $\theta_{13}$
at about $1.5\,\sigma$~\cite{LisiHint08,theta13review}. 

In preparation for the Double Chooz reactor experiment,
we have re-evaluated the specific reactor antineutrino flux 
($\nu$/fission), improving the electron to antineutrino data
conversion~\cite{LM}. The method relies on detailed knowledge of
the decays of thousands of fission products, while the previous
conversion procedure used a phenomenological model based on
30~effective beta branches.  
Both methods are constrained by the well-measured ILL spectrum 
of fission induced electrons that accompanies 
the antineutrinos~\cite{SchreckU5_80,SchreckU5_85,SchreckU5Pu9,SchreckPu9Pu1}.

The new calculation results in a small increase in the flux of~3.5\%.
Although this increase has no significant effect on KamLAND's
solar parameter results, 
when combined with the previously reported small deficits at
nearer distances, it results in a larger average deficit of~5.7\%, 
at $0.943\pm 0.023$. We call it the reactor antineutrino anomaly,  
significant at the level of 98.6\%~C.L. This deficit is consistent with
being independent from the distance to the reactor core at distances~$\gtrsim 15$~meters,
the distance of the Bugey-3 experiment~\cite{Bugey3}.

If this deficit is due to neutrino mixing, it could be explained by
an energy-independent suppression of the $\nuebar$ flux at
distances~$\gtrsim 15$~meters. This requires a neutrino
with a $|\Delta m_{\rm new}^2|\gtrsim 1$~eV$^2$.
The mixing amplitude with the $\nu_{e}$ must be $\sin^2(2\theta_{\rm new}) \sim 0.115$. 
The required $|\Delta m_{\rm new}^2|$ is significantly larger than those required
by solar and atmospheric experiments. 
This suggests, if the neutrino mixing hypothesis is the correct
explanation, the existence of a fourth neutrino, beyond the standard model. 
We would like to stress here that other explanations are also possible, such as a
correlated artifact in the experiments, or an erroneous prediction of
the antineutrino flux from nuclear reactor cores.

We begin by discussing the use of new antineutrino spectra and provide 
a revised value of the predicted cross section per fission, in
Section~\ref{newsigmaf}.
We then revisit the short baseline oscillation results ($<100$~m)
and CHOOZ in Section \ref{sblexp}, revealing a reactor
antineutrino anomaly, which is discussed in Section \ref{anomaly}. 
The compatibility of our results with the existence of a fourth
non-standard neutrino state is discussed in Section \ref{newphysics}.
The CHOOZ and KamLAND sensitivities to the $\theta_{13}$ mixing angle
are revisited in Sections \ref{chooz} and \ref{kamland}. 
Their combination is discussed in Section \ref{combined}.
Two experimental programs for testing the anomaly are discussed in
Section \ref{testanomaly}. We conclude in Section \ref{conclusion}.

\section{New Predicted Cross Section per Fission}
\label{newsigmaf}

Fission reactors release about $10^{20}\,\nuebar\;{\rm GW}^{-1}{\rm s}^{-1}$, which
mainly come from the beta decays of the fission products of
 $^{235}$U, $^{238}$U, $^{239}$Pu, and $^{241}$Pu. 
The emitted antineutrino spectrum is then given by:
\begin{equation}
S_{\rm tot} (E_\nu) =  \sum_{k} f_k S_{k}(E_\nu), 
\end{equation}
where $f_k$ refers to the contribution of the main fissile nuclei to
the total number of fissions of the k$^{\rm th}$ branch, 
and $S_k$ to their corresponding neutrino spectrum per fission.

For the last 25 years the $\bar{\nu}_e$ spectra have been estimated
from measurements of the total electron spectra associated with the beta
decays of all fission products of $^{235}$U, $^{239}$Pu, and
$^{241}$Pu. Thin target foils of these isotopes were irradiated with
thermal neutrons at the ILL
reactor~\cite{SchreckU5_85,SchreckU5Pu9,SchreckPu9Pu1}. 
The measured spectra then had to be converted from electron to
antineutrino spectra invoking a set of
30 effective beta-branches, adjusted to reproduce the total electron
spectrum~\cite{VogelXsec}. 

Recently we revisited the conversion procedure with a novel
mixed-approach combining the accurate reference of the ILL electron
spectra with the physical distribution of beta branches of all
fission products provided by the nuclear databases~\cite{LM}. 
This new approach provided a better handle on the systematic errors of the
conversion. Although it did not reduce the final error budget, it led
to a systematic shift of about 3\% in the normalization 
of $^{235}$U, $^{239}$Pu, and $^{241}$Pu antineutrino fluxes,
respectively. This normalization shift has been attributed to two
main systematic effects in the original conversion of the ILL
electron data. At low energy ($E_{\nu}<4$~MeV) the implementation of Coulomb
and weak magnetism corrections to the Fermi theory in the new approach
turned out to deviate from the effective linear correction (\mbox{$0.65
\times(E_{\nu}-4$~MeV}) in~\%) used in the previous work. At high energy
($E_{\nu}>4$~MeV), the
converted antineutrino spectra become very sensitive to the knowledge
of the charge $Z$ of the nuclei contributing to the total
spectrum. In the previous approach, only the mean dependence of $Z$
versus the end-point of the effective beta-branches had been used while
in the new conversion we had access to the complete distribution,
nucleus by nucleus. These two effects could be numerically studied and
confirmed on various independent sets of beta-branches.

Because $^{238}$U nuclei undergo fission with fast neutrons, the associated
electron spectrum could not be measured in the thermal neutron flux of the ILL
reactor. Therefore the {\it ab initio} summation of the $\bar{\nu}_e$  from
all possible beta decays of fission products was performed to predict
the neutrino spectrum~\cite{CalcU8}. In Ref.~\cite{LM} we provided a new prediction
with an estimated relative uncertainty of the order of 15\% in the 2-8~MeV range.
This uncertainty of {\it ab initio}
calculations is still too large to be generalized to all isotopes but
it is sufficiently accurate in the case of $^{238}$U, which contributes to less than
10\% of the total fission rate for all reactors considered in this
work. An ongoing measurement at the FRM II reactor in Garching will soon provide
experimental constraints~\cite{Nils}.

When predicting the antineutrino spectrum of a reactor $S_{\rm tot}(E_\nu)$, a weighted sum
of the four antineutrino spectra $S_{k}(E_\nu)$ is computed according to the considered fuel
composition, which can be different in each experiment. The object of this article is to analyze the
impact of the above-mentioned $\sim 3$~percent shift on past, 
present, and future experiments.

Generally reactor neutrino oscillation experiments search for the
reaction:
\begin{equation}
\nuebar+p\rightarrow e^{+}+n ,
\label{idb}
\end{equation}
where an electron antineutrino interacts with a free proton in a 
detector, often filled with scintillator. The reaction cross section
can be precisely computed with the \mbox{V-A}~theory of weak interaction~\cite{FAY85}:
\begin{equation}
\sigma_{\rm{V-A}} (E_e)= \kappa  p_e E_e (1+\delta_{\rm rec}+\delta_{\rm wm}+\delta_{\rm rad}),
\label{sigVA}
\end{equation}
$p_e$ and $E_e$ being the momentum
and energy of the positron, and the $\delta$ being the 
energy dependent recoil ($\delta_{\rm rec}$), weak magnetism
($\delta_{\rm wm}$), and radiative ($\delta_{\rm rad}$) corrections.
On the one hand, the prefactor $\kappa$ can be written 
\begin{equation}
\kappa = \frac{G^2_F cos^2\theta_C}{\pi}
(1+\Delta^R_{\rm inner})(1+3\lambda^2),
\label{kappa}
\end{equation}
where $G_F$ is the Fermi constant, $\theta_C$ the Cabibbo angle,
$\Delta^R_{\rm inner}=0.024$ the inner radiative corrections taken from
Ref.~\cite{Wil94}, and $\lambda=1.2694$ the form factor ratio of the axial to vector
coupling constant. Parameters are taken by default to
their latest {\rm PDG} average values~\cite{pdg}. 
Using this parameterization Ref.~\cite{VB99} obtained a reference
value of $\kappa=0.952~10^{-43}$~cm$^2$, using $\lambda=1.2670$
close to the average value of the year 1999~\cite{PDG99}. This value must now be
updated to 1.2694~\cite{pdg}, leading to $\kappa=0.955~10^{-43}$~cm$^2$.

On the other hand, Eq.~\ref{sigVA} can be normalized to the $\beta$-decay of the free
neutrons and $\kappa$ can be written as:
\begin{equation}
\kappa = \frac{2 \pi^2}{m_e^5 f^R \tau_n},
\end{equation}
$\tau_n$ being the neutron mean lifetime, and  $f^R=1.71465(15)$ the
phase-space factor for beta-decay of the free neutron taken from
Ref.~\cite{Wil82}, including outer radiative corrections.
Over the last 15~years the neutron mean lifetime has evolved, from 926~s
(value used in the ILL experiment~\cite{ILL81}) to its current {\rm PDG} 
average value of 885.7~s~\cite{pdg}. Based on this parameterization
the current prefactor $\kappa$ is $0.956~10^{-43}$~cm$^2$. This is
our default value in this publication. For the computation of the
differential cross section we used~\cite{FAY85}. Our results agree within 0.1\% with the
results published in Ref.~\cite{VB99} that supersede those of Ref.~\cite{VogelXsec}.

It is worth mentioning that there is an ongoing controversy about the
world average neutron lifetime, with Ref.~\cite{Serebrov} finding
$\tau_n=878$~s,  and making the world average `suspect'.
A lower value has also been found by the MAMBO-2 group~\cite{Mambo}.
The new world average should then evolve and settle to 881.4(1.4)~s in~2011
(Ref.~\cite{Mambo} and private communication from K.~Schreckenbach).

We note here that the average value of $\lambda$ may depend on neutron
lifetime measurements. However other experiments studying angular
correlations between the neutron spin and the emitted electron provide
independent measurements of $\lambda$.  Recently Ref.~\cite{Liu} finds~1.27590
 and  Ref.~\cite{Abele} favors $\lambda=1.2750(9)$. 
A value of $\lambda=1.274$ would be consistent with the latest neutron life
time measurements, to be averaged to $\tau_n$=881.4~s~\cite{Mambo}.
Thus the latest values of $\lambda$ and $\tau_n$ point to a
reevaluation of the prefactor $\kappa$ to $0.961~10^{-43}$~cm$^2$.
With this change, the cross section would increase by 0.5\% compared
to the calculations in this work. 

The outgoing positron and incoming antineutrino energies are related
by 
\begin{equation}
E_\nu  =  E_e +\Delta +
\frac{E_e(E_e+\Delta)}{M}+\frac{1}{2}\frac{(\Delta^2-m_e^2)}{M},
\end{equation}
where  $\Delta=M_n-M_p$~\cite{VogelXsec, FAY85}.
The prediction of the cross section per fission is defined as: 
\begin{equation}
\label{sigmaexp}
\sigma^{\rm pred}_{f}   =  \int_{0}^{\infty}  S_{\rm tot} (E_\nu) \sigma _{\rm{V-A}}
(E_\nu) dE_\nu =  \sum_{k} f_k \sigma^{\rm pred}_{f,k} , 
\end{equation}
where $S_{\rm tot}$ is the model dependent reactor neutrino spectrum 
for a given average fuel composition ($f_k$) and $\sigma_{\rm{V-A}}$ is the theoretical
cross section of reaction (\ref{idb}). The $\sigma^{\rm pred}_{f,k}$ are
the predicted cross sections for each fissile isotope.
Experiments at baselines below $100$~m reported either the
ratios of the measured to predicted cross section per fission, or
the ratios (R) of the observed event rate to the predicted rate.

Accounting for new reactor antineutrino spectra~\cite{LM} the
normalization of predicted antineutrino rates, $\sigma^{\rm pred}_{f,k}$, is shifted by 
+2.5\%, +3.1\%, +3.7\%, +9.8\% for  k=$^{235}$U, $^{239}$Pu,
$^{241}$Pu, and $^{238}$U respectively (see Table~\ref{tab:crosssec}). 
In the case of $^{238}$U the completeness of nuclear databases over the years 
largely explains the +9.8\% shift from the reference computations~\cite{CalcU8}.
The new predicted cross section for any fuel composition
can be computed from Eq.~(\ref{sigmaexp}).
By default our new computation takes into account the so-called
off-equilibrium correction of the antineutrino fluxes 
(increase in fluxes caused by the decay of long-lived
fission products). 

\section{Impact on past experimental results}
\label{sblexp}
In the eighties and nineties, experiments were performed at a few tens 
of meters from nuclear reactor cores at ILL, Goesgen, Rovno, Krasnoyarsk, 
Bugey (so called~3 and~4) and Savannah River~\cite{ILL81,Goesgen,Rovno88,
Rovno91,Krasnoyarsk_87,Krasnoyarsk_94,Bugey3,Bugey4, SRP96}. 
Following these pioneering results middle- and long-baseline experiments were
performed at CHOOZ~\cite{Chooz} and KamLAND~\cite{Kamland} 
in the late nineties and after. In this section we revised each ratio 
of observed event rate to predicted rate. The observed event rates and
their associated errors are unchanged. The predicted rates are reevaluated
separately in each experimental case.

\subsection{The Bugey-4 integral measurement}
\label{bugey}

The Bugey-4 experiment~\cite{Bugey4} performed the most precise 
measurement of the inverse beta decay cross section, 
obtaining $\sigma^{\rm Bugey}_{f}$=5.752$\pm0.081$ in units of 10$^{-43}$~cm$^2$/fission. 
This experimental result was compared to the predicted cross section
per fission, $\sigma^{\rm pred,old}_{f}$. 

Using Ref.~\cite{Bugey4} inputs, and the former converted neutrino
spectra from~\cite{SchreckU5_85, Vogel81, Kopeikin88}
(except for the $^{238}$U neutrino spectrum taken from~\cite{LM}), 
we get $\sigma^{\rm pred,old}_{f}$=5.850~10$^{-43}$~cm$^2$/fission, in good
agreement with Bugey-4's estimation (see Table~\ref{tab:crosssec}). 
We used the average fuel composition of Ref.~\cite{Bugey4}, $^{235}$U=53.8\%,
$^{239}$Pu=32.8\%, Pu$^{241}$=5.6\% and U$^{238}$=7.8\%, in fractions
of fissions per isotope.
We note here that the published Bugey-4 cross section calculation does not account for 
the contribution of long-lived fission products (off-equilibrium
effects). The reference electron spectra used for the determination of $S_{\rm tot}$
were acquired after an irradiation time in the ILL reactor flux of less
than 1.5~day for all isotopes. But in the Bugey experiments at a commercial PWR the 
irradiation time scale was typically about one year. $S_{\rm tot}$ should thus be 
corrected for the accumulation of long-lived fission 
products in the low energy part of the spectrum. 
We had to turn off these effects in our computation to recover
the $\sigma^{\rm pred,old}_{f}$~value. 
Including these effects in our calculations would lead to a +1.0\% discrepancy. 
An over or under-estimation of the irradiation time by 100~days
would lead to a systematic error on the off-equilibrium correction below 0.1\%.

Computing the ratio of observed versus expected events 
\mbox{$R_{\rm Bugey}$=$\sigma^{\rm Bugey}_{f}$/$\sigma^{\rm pred}_{f}$}, Bugey-4 
concluded that there was good agreement with the V-A theory of weak 
interactions, and that the neutrino flux emitted by PWR reactors is sufficiently
understood to be computed using the parameters of~\cite{SchreckU5_85, Vogel81,
Kopeikin88}.

Applying the new reactor antineutrino spectra we obtain a 
new value of the cross section per fission of
\mbox{$\sigma^{\rm pred, new}_{f}=6.102\pm 0.163$}
in units of 10$^{-43}$~cm$^2$/fission.  
We thus revised the ratio~\mbox{$R_{\rm Bugey-4}$=$0.943\pm 0.013\,({\rm stat}+{\rm syst})\pm0.025\,(S_{\rm tot})$},
which is now 2.0~standard deviations from the expectation without neutrino oscillations.
This creates a tension between the measurement at Bugey-4 and the new predicted cross section
per fission. In the next sections we will see that the other reactor neutrino
experimental results confirm this anomaly. 

\begin{table}[t]
\begin{center}
\medskip
\begin{tabular}{c|c|c}
\hline\hline
                              & old~\cite{Bugey4}       & new  \\
\hline
$\sigma^{\rm pred}_{f,^{235}{\rm U}}$    &  6.39$\pm$1.9\%  & 6.61$\pm$2.11\%  \\
$\sigma^{\rm pred}_{f,^{239}{\rm Pu}}$   &  4.19$\pm$2.4\%  & 4.34$\pm$2.45\%   \\
$\sigma^{\rm pred}_{f,^{238}{\rm U}}$    &  9.21$\pm$10\%   & 10.10$\pm$8.15\% \\
$\sigma^{\rm pred}_{f,^{241}{\rm Pu}}$   &  5.73$\pm$2.1\%  & 5.97$\pm$2.15\%   \\
\hline
$\sigma^{\rm pred}_{f}$                  & 5.824$\pm$2.7\% & 6.102$\pm$2.7\% \\
\hline
$\sigma^{\rm Bugey}_{f}$                 & \multicolumn{2}{c}{5.752$\pm$1.4\% \cite{Bugey4}} \\
$\sigma^{\rm Bugey}_{f}$/$\sigma^{\rm pred}_{f}$ & 0.987$\pm$1.4\%$\pm$2.7\% & 0.943$\pm$1.4\%$\pm$2.7\%  \\
\hline\hline
\end{tabular}
\caption{\label{tab:crosssec} Individual cross sections per fission
per fissile isotope, $\sigma^{\rm pred}_{f,k}$. 
The slight differences in the ratios are caused by differences in off-equilibrium effects. 
Results of the comparison of $\sigma^{\rm Bugey}_{f}$ at Bugey-4 in units of 
10$^{-43}$~cm$^2$/fission with the former and present
prediction, $\sigma^{\rm pred}_{f}$, are shown. }
\end{center}
\end{table}

\subsection{The ILL neutrino experiment}
\label{ill}
In the early eighties the ILL experiment measured the antineutrino
induced positron spectrum at a distance of 8.76~m from the core of the Laue-Langevin
fission reactor. Its fuel is highly enriched uranium (93\% $^{235}$U),
simplifying the computation of the predicted antineutrino
spectrum rate and shape. 
In Ref.~\cite{ILL81} the ILL experiment reported a ratio of measured to
predicted event rates of $R_{\rm ILL,80}$=$0.955\pm0.035\,({\rm stat})\pm0.11\,({\rm syst}+S_{\rm tot})$. 

In 1995 this result was revised by part of the ILL collaboration~\cite{ILL95}. 
The 1980 reactor electron spectrum for $^{235}$U of Ref.~\cite{SchreckU5_80} 
was re-measured in 1984~\cite{SchreckU5_85} by the same authors as
\cite{SchreckU5_80}. The neutron mean lifetime was corrected from
926~s to 889~s, increasing the predicted cross section by 4\%. 
Moreover in 1990 it was announced that the operating power of the
ILL reactor had been incorrectly reported at the time of the neutrino
experiment, by +9.5\%. This reanalysis led to 
\mbox{$R_{\rm ILL,95}$=$0.832\pm0.029\,({\rm stat})\pm0.0738\,({\rm syst}+S_{\rm tot})$}, excluding the
no-oscillation case at $2\,\sigma$. 

According to the new spectra of Ref.~\cite{LM} the antineutrino rate 
is increased by $\sim$3.5\% (see Table \ref{tab:other}). 
A slight neutron mean lifetime correction leads to an additional 
+0.37\% shift. The new ratio is thus
\mbox{$R_{\rm ILL,new}$=$0.802\pm0.028\,({\rm stat})\pm0.071\,({\rm syst}+S_{\rm tot})$},
significantly deviating from its expected value.

\subsection{Bugey-3, Goesgen, Krasnoyarsk, Rovno, SRP}
\label{others}

We now study the impact of the new antineutrino spectra on experiments at 
Bugey (called Bugey-3), Goesgen, Rovno, and Krasnoyarsk~\cite{Goesgen,Rovno88,
Rovno91,Krasnoyarsk_87,Krasnoyarsk_94,Bugey3}, 
which measured the reactor antineutrino rate at short distances, between 15~m and 95~m,
but less accurately than Bugey-4~\cite{Bugey4}.

Accounting for new reactor antineutrino spectra~\cite{LM} the Bugey-3~\cite{Bugey3}
ratios of observed versus expected events between~1 and 6~MeV 
decrease by~3.7\%. However, we have to apply a further
correction to account for off-equilibrium effects. 
Assuming 300~days of irradiation this leads to an additional increase by~+1.0\%.
The Bugey-3 ratios become:
\mbox{$0.946 \pm0.004\,({\rm stat}) \pm0.048\,({\rm syst}+S_{\rm tot})$} at
15~m,
\mbox{$0.952\pm0.01\,({\rm stat}) \pm0.048\,({\rm syst}+S_{\rm tot})$}
at 40~m and
\mbox{$0.876 \pm0.126\,({\rm stat}) \pm 0.048\,({\rm syst}+S_{\rm tot})$}
at 95~m. Note that uncertainties on $S_{\rm tot}$ are
included in the errors quoted by the Bugey-3 collaboration.

A similar analysis is performed with the Goesgen~\cite{Goesgen}
results. The new Goesgen ratios shift to 
\mbox{$0.966 \pm 0.017\,({\rm stat}) \pm 0.060\,({\rm syst}+S_{\rm tot})$}
at 38~m,
\mbox{$0.991\pm0.019\,({\rm stat}) \pm 0.062\,({\rm syst}+S_{\rm tot})$}
at 45~m and
\mbox{$0.924 \pm0.033\,({\rm stat}) \pm0.062\,({\rm syst}+S_{\rm tot})$} at 65~m.

In the eighties a series of experiments were performed at the Rovno nuclear
power station in the Soviet Union.

In~1988 the collaboration published measurements at 18~m and 25~m from
the reactor core~\cite{Rovno88}. Five measurements
were performed with two different detectors: an integral detector to measure the absolute cross
section per fission of inverse $\beta$-decay (labeled~1I and~2I), and a scintillator
spectrometer to measure both the absolute and differential cross
section per fission (labeled~1S, 2S, and~3S). The neutron lifetime was taken to be 898.8~s.  We
accounted for the average fuel composition for each run which was published in~\cite{Rovno88}. As an
indication, the average fuel composition over the five measurements is
$^{235}$U=59.6\%, $^{239}$Pu=28.3\%, $^{241}$Pu=4.6\%  and
U$^{238}$=7.5\%.The five results of the observed over predicted event
ratios are reported in table \ref{tab:other}. At 18~m the average ratio
is shifted from 
\mbox{$R_{\rm Rovno88,18m,new}$=$0.995\pm0.060\,({\rm stat}+{\rm syst}+S_{\rm tot})$}
to
\mbox{$R_{\rm Rovno88,18m,new}$=$0.944\pm0.057\,({\rm stat}+{\rm syst}+S_{\rm tot})$}.

In~1991 the Rovno integral experiment~\cite{Rovno91} published a
cross section per fission of $\sigma^{\rm Rovno91}_{f}=5.85\pm0.17$ in
units of 10$^{-43}$~cm$^2$/fission, 18~m away from a nuclear core
with an average fuel composition of 
$^{235}$U=61.4\%, $^{239}$Pu=27.4\%, $^{241}$Pu=3.8\%  and
U$^{238}$=7.4\%. They predicted the cross section 
$\sigma^{\rm pred,old}_{f,{\rm Rovno91}}=5.94\pm0.16$ in units
of~10$^{-43}$~cm$^2$/fission, and thus obtained the ratio
\mbox{$R_{\rm Rovno91,old}$=$0.985\pm0.037\,({\rm stat}+{\rm syst}+S_{\rm tot})$}.
We recomputed the cross section per fission according to the new
antineutrino spectra and found 
\mbox{$\sigma^{\rm pred,new}_{f,{\rm Rovno91}}=6.223\pm0.17$}. 
The new ratio is thus revised to
\mbox{$R_{\rm Rovno91,new}$=$0.940\pm0.036\,({\rm stat}+{\rm syst}+S_{\rm tot})$}.
We note that the correction of the neutron mean lifetime contributes~1.8\%
to the shift of the ratio. 

In~1984 a neutrino experiment operated at the Krasnoyarsk
reactors~\cite{Krasnoyarsk_87}, which have an almost pure $^{235}$U fuel
composition leading to an antineutrino spectrum within~1\% of
pure $^{235}$U, and operate over 50~day cycles. They measured 
the cross section per fission at two distances,
\mbox{$\sigma^{\rm Krasno,33m}_{f}=6.19\pm0.36$} at 33~m and 
\mbox{$\sigma^{\rm Krasno,92m}_{f}=6.30\pm1.28$} at 92~m, in
units of~10$^{-43}$~cm$^2$/fission. They compared it to the predicted
cross section of $6.11\pm0.21$~10$^{-43}$~cm$^2$/fission, based on the 
Ref.~\cite{SchreckU5_80} $^{235}$U measurement instead of Ref.~\cite{SchreckU5_85}.
Correcting the neutron mean lifetime and using the new antineutrino
spectra we obtain a predicted cross section of 6.61~10$^{-43}$~cm$^2$/fission,
assuming a pure $^{235}$U spectrum. This leads to the ratios 
\mbox{$R_{\rm Krasno,33m}$=$0.936\pm0.054\,({\rm stat}+{\rm syst}+S_{\rm tot})$} and 
\mbox{$R_{\rm Krasno,92m}$=$0.953\pm0.195\,({\rm stat}+{\rm syst}+S_{\rm tot})$}, at 33~m
and 92~m, respectively.
In~1994 two other measurements were performed 57~m from
the Krasnoyarsk reactors~\cite{Krasnoyarsk_94}. 
They measured \mbox{$\sigma^{\rm Krasno,57m}_{f}=6.26\pm0.26$} at 57~m, and
compared it to their predicted cross section of $6.33\pm0.19$~10$^{-43}$~cm$^2$/fission,
based on Ref.~\cite{SchreckU5_85} and in
agreement with our reevaluation using previous reference antineutrino
spectra. Using the new values of Ref.~\cite{LM} we revise the ratio 
\mbox{$R_{\rm Krasno,57m}$=$0.947\pm0.047\,({\rm stat}+{\rm syst}+S_{\rm tot})$}.

From the neutrino pioneering experiments led by F.~Reines and C.~Cowan~\cite{Reines56}
to the nineties, a series of reactor antineutrino
measurements were performed at the Savannah River Plant (SRP), a U.S
production facility for tritium and plutonium. For neutrino energies
between 2~and 8~MeV the spectrum difference between SRP and a similar
core with pure $^{235}$U fuel was estimated to be less than~0.5\%. 
We make use of the latest results
published in Ref.~\cite{SRP96}. Measurements were reported at two
different baselines, 18~m and 24~m. The new SRP ratios are reevaluated to
\mbox{$0.953 \pm 0.006\,({\rm stat}) \pm 0.0353\,({\rm syst}+S_{\rm tot})$} and
\mbox{$1.019 \pm 0.010\,({\rm stat}) \pm 0.0377\,({\rm syst}+S_{\rm tot})$},
respectively.

\begin{table*}[t]
\begin{center}
\medskip
\begin{tabular}{c|c|c|c|c|c|c|c|c|c|c|c|c}
  \hline\hline
  \# & result   & Det. type & $\tau_n$ (s)& $^{235}$U & $^{239}$Pu &$^{238}$U & $^{241}$Pu& old  & new  & err(\%) & corr(\%) & L(m) \\
  \hline
  1&Bugey-4     & $^3$He+H$_2$O & 888.7 & 0.538 & 0.328 & 0.078 & 0.056 & 0.987 & 0.942 & 3.0  & 3.0 & 15\\
  2&ROVNO91     & $^3$He+H$_2$O & 888.6 & 0.614 & 0.274 & 0.074 & 0.038 & 0.985 & 0.940 & 3.9  & 3.0 & 18\\
  \hline
  3&Bugey-3-I   & $^6$Li-LS     & 889   & 0.538 & 0.328 & 0.078 & 0.056 & 0.988 & 0.946 & 4.8  & 4.8 & 15\\ 
  4&Bugey-3-II  & $^6$Li-LS     & 889   & 0.538 & 0.328 & 0.078 & 0.056 & 0.994 & 0.952 & 4.9  & 4.8 & 40\\ 
  5&Bugey-3-III & $^6$Li-LS     & 889   & 0.538 & 0.328 & 0.078 & 0.056 & 0.915 & 0.876 & 14.1 & 4.8 & 95\\
  \hline
  6&Goesgen-I   & $^3$He+LS     & 897   & 0.620 & 0.274 & 0.074 & 0.042 & 1.018 & 0.966 & 6.5  & 6.0 & 38\\
  7&Goesgen-II  & $^3$He+LS     & 897   & 0.584 & 0.298 & 0.068 & 0.050 & 1.045 & 0.992 & 6.5  & 6.0 & 45\\
  8&Goesgen-II  & $^3$He+LS     & 897   & 0.543 & 0.329 & 0.070 & 0.058 & 0.975 & 0.925 & 7.6  & 6.0 & 65\\
  9&ILL         & $^3$He+LS     & 889   & $\simeq1$ &--- &---   &---    & 0.832 & 0.802 & 9.5  & 6.0 &  9\\
  \hline
  10&Krasn. I   & $^3$He+PE     & 899   & $\simeq1$ &--- &---   &---    & 1.013 & 0.936 & 5.8  & 4.9 & 33\\
  11&Krasn. II  & $^3$He+PE     & 899   & $\simeq1$ &--- &---   &---    & 1.031 & 0.953 & 20.3 & 4.9 & 92\\
  12&Krasn. III & $^3$He+PE     & 899   & $\simeq1$ &--- &---   &---    & 0.989 & 0.947 & 4.9  & 4.9 & 57\\
\hline
  13&SRP I      & Gd-LS         & 887   & $\simeq1$ &--- &---   &---    & 0.987 & 0.952 & 3.7  & 3.7 & 18\\
  14&SRP II     & Gd-LS         & 887   & $\simeq1$ &--- &---   &---    & 1.055 & 1.018 & 3.8  & 3.7 & 24\\
\hline
  15&ROVNO88-1I & $^3$He+PE     & 898.8 & 0.607 & 0.277 & 0.074 & 0.042 & 0.969 & 0.917 & 6.9  & 6.9 & 18\\ 
  16&ROVNO88-2I & $^3$He+PE     & 898.8 & 0.603 & 0.276 & 0.076 & 0.045 & 1.001 & 0.948 & 6.9  & 6.9 & 18\\
  17&ROVNO88-1S & Gd-LS         & 898.8 & 0.606 & 0.277 & 0.074 & 0.043 & 1.026 & 0.972 & 7.8  & 7.2 & 18\\
  18&ROVNO88-2S & Gd-LS         & 898.8 & 0.557 & 0.313 & 0.076 & 0.054 & 1.013 & 0.959 & 7.8  & 7.2 & 25\\
  19&ROVNO88-3S & Gd-LS         & 898.8 & 0.606 & 0.274 & 0.074 & 0.046 & 0.990 & 0.938 & 7.2  & 7.2 & 18\\
%\hline
%20&PaloVerde   & Gd-LS         & 885.7 &       &       &       &       & 1.011 & 0.975 & 6.1  & --- & 820\\
%\hline
%21&CHOOZ       & Gd-LS         & 886.7 &       &       &       &       & 1.010 & 0.961 & 4.3  & --- & 1050\\
\hline\hline
\end{tabular}
\caption{\label{tab:other} $N_{\rm obs}/N_{\rm pred}$ ratios based on
  old and new spectra. Off-equilibrium corrections have been applied when justified.  
  The err column is the total error published by the
  collaborations including the error on $S_{\rm tot}$, 
  the  corr column is the part of the error correlated among
  multiple-baseline experiments, or experiments using the same detector.
  This table is used to construct the covariance matrix used in Eq.~\ref{chi2cov}.
}
\end{center}
\end{table*}

\subsection{CHOOZ and Palo Verde}
\label{choozratio}

Based on the good agreement between $\sigma^{\rm pred,old}_{f}$ and
$\sigma^{\rm Bugey}_{f}$ obtained at Bugey-4~\cite{Bugey4}, 
the CHOOZ experiment~\cite{Chooz} decided to use the total cross 
section per fission measured at Bugey-4,
correcting a posteriori for the different averaged fuel composition 
according to:
\begin{equation}
\sigma^{\rm Chooz}_{f} = \sigma^{\rm Bugey}_{f} + \sum_{k} (f_k^{\rm Chooz}-f_k^{\rm Bugey})\sigma^{\rm pred,old}_{f,k} , 
\end{equation}
where $f_k^{\rm Chooz}$ and $f_k^{\rm Bugey}$ are the contributions of the $k^{\mathrm{th}}$
isotope to the total amount of fissions at the CHOOZ and 
Bugey-4 experiments~\cite{Nicolo}.
This explicitly means that the expected number of
events was computed using $\sigma^{\rm Bugey}_{f}$ rather than $\sigma^{\rm pred,old}_{f}$,
thus absorbing a~-1.3\% difference on the overall normalization (see
Table \ref{tab:crosssec}).
This also has the effect of reducing the error on the
neutrino detection rate from~2.7\% to~1.6\% 
(including the uncertainty on the fission contributions $f_k$). 
Accounting for an uncertainty on off-equilibrium
effects, CHOOZ quoted a final neutrino spectrum error of~1.9\%.
As shown above, the new values from Ref.~\cite{LM} lead to an increase
of $\sigma^{\rm pred,new}_{f}$ = 1.048~$\sigma^{\rm pred,old}_{f}$. With
this sizeable discrepancy between measured and computed cross sections,
the CHOOZ experiment cannot rely anymore on the effective cross
section per fission measured at Bugey-4, assuming no-oscillation at
baselines of less than a few tens of meters. 
We thus revise the ratio to
\mbox{$R_{\rm Chooz}$=$0.961\pm0.027\,({\rm stat})\pm0.032\,({\rm syst}+S_{\rm tot})$}
(see Table~\ref{tab:other}), where the~3.3\% systematic
error now includes the~2.7\% uncertainty on~$\sigma^{\rm pred,new}_{f}$
(See Appendix~\ref{errprop}).

A crude analysis of the impact of Ref.~\cite{LM} on Palo Verde 
data~\cite{Paloverde} leads to the modification of the average ratio
of detected versus expected $\nuebar$ by roughly~$-3.5\%$, leading 
to \mbox{$R_{\rm PaloVerde}=0.975\pm0.023\,({\rm stat})\pm0.055\,({\rm syst}+S_{\rm tot})$}.
We also note that Palo Verde's uncertainty would increase from~5.3\% in Ref.~\cite{Paloverde}
to~5.6\% according to our prescription.

\section{The reactor antineutrino anomaly}
\label{anomaly}

\subsection{Rate information}
\label{anorate} 

Up to now we independently studied the results of the main reactor
neutrino experiments using a new value of the cross section
per fission, $\sigma^{\rm pred,new}_{f}$. 
The ratios of observed event rates to predicted event rates,
\mbox{$R=N_{\rm obs}/N_{\rm pred}$}, are summarized in Table~\ref{tab:other}.
We observe a general systematic shift more or less significantly below unity. 
These reevaluations unveil a new {\it reactor antineutrino anomaly}, 
clearly illustrated in Figure~\ref{f:anomalyfit}, but still to be
explained. 
In order to quantify the statistical significance of the anomaly we
can compute the weighted average of the ratios of expected over
predicted rates, for all short baseline reactor
neutrino experiments (including their possible correlations). 

\begin{figure}[!h]
\begin{center}
\includegraphics[scale=0.5]{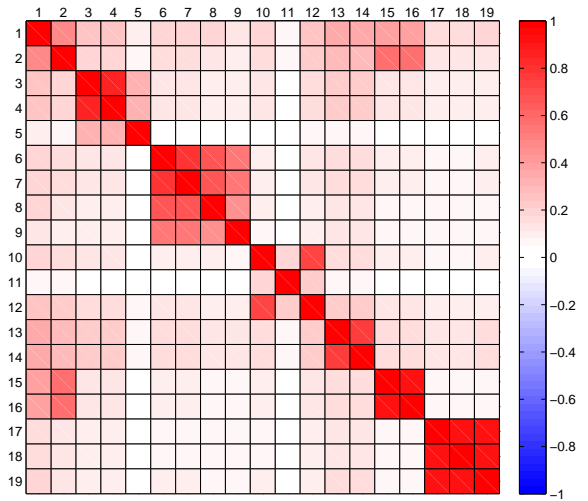}
\caption{\label{covmat} Correlation matrix of 19~measurements
  of reactor neutrino experiments operating at short
  baselines. Experiment labels are given in Table~\ref{tab:other}.}
\end{center}
\end{figure}

We consider the following experimental rate information:
Bugey-4 and Rovno91, the three Bugey-3
experiments, the three Goesgen experiments and the ILL experiment,
the three Krasnoyarsk experiments, the two Savannah River results
(SRP), and the five Rovno88 experiments.
$\overrightarrow{\text{R}}$~is the corresponding vector of 19~ratios
of observed to predicted event rates.
We assume a~2.0\% systematic uncertainty fully correlated among all
19~ratios. This choice is motivated by the common
normalization uncertainty of the corresponding beta-spectra measured
in~\cite{SchreckU5_80,SchreckU5_85,SchreckU5Pu9,SchreckPu9Pu1}.
We considered the ratios and relative errors gathered in Table~\ref{tab:other}.
In order to account for the potential experimental correlations, we fully
correlated the experimental errors of Bugey-4 and Rovno91, of the three
Goesgen and the ILL experiments, the three Krasnoyarsk
experiments, the five Rovno88 experiments, and the two SRP results.
We also fully correlated the Rovno88 (1I~and~2I) results with
Rovno91, and we added an arbitrary~50\% correlation between the Rovno88
(1I~and~2I) and the Bugey-4 measurement. We motivated these latest
correlations by the use of similar or identical integral detectors.
We stress here that in this publication we only used the error budget
published by the collaborations, without any change.
We then obtain the covariance matrix W of the ratios. In Figure~\ref{covmat}
we show the corresponding correlation matrix with labels detailed in
Table~\ref{tab:other}.

All the above-mentioned experiments published ratios of measured to expected event rates.
While the rates themselves can be considered to follow Gaussian distributions, the
ratio of two Gaussian variables does not. This may lead to overestimating the statistical
significance of the deviation of the ratios from unity, as was pointed out by James
in response to a paper by Reines {\emph et al.}~\cite{reinesjames}.
In order to check our results, we performed simple Monte-Carlo simulations: each experiment's
observed and expected rates were simulated within their quoted errors, including
the correlations between experiments shown in the covariance matrix W.

We then calculated the best-fit average ratio~$r_{\mathrm{best}}$
for each simple Monte-Carlo experiment by minimizing the~$\chi^2$ function
with respect to $r$:
\begin{equation}
\left( 
r-\overrightarrow{\text{R}}
\right)^T 
W^{-1}
\left( 
r-\overrightarrow{\text{R}}
\right), 
\label{chi2cov}
\end{equation}
and obtained the distribution of~$r_{\mathrm{best}}$ 
(we note $\mu$ its mean in the following paragraph).
We found that it is almost Gaussian, but with slightly longer tails,
which we decided to take into account in our
calculations (in contours that appear later in this article we enlarged the
error bars).
If we only consider experiments with baselines $<$~100~m to get rid of
a possible ($\theta_{13}$, $\Delta m_{31}^2$) driven oscillation
effect at Palo Verde or CHOOZ, with
the old antineutrino spectra the mean ratio is $\mu$=0.976$\pm$0.024,
and the fraction of simple Monte-Carlo experiments with
$r\ge 1$ is 17.1\% ($-0.95\,\sigma$ from expectation).
With the new antineutrino spectra, we obtain $\mu$=0.943$\pm$0.023, and the fraction
of simple Monte-Carlo experiments with $r \ge 1$ is 1.3\%, corresponding to a
$-2.2\,\sigma$ effect (while a simple calculation assuming normality
would lead to $-2.4\,\sigma$). Clearly the new spectra induce a statistically
significant deviation from the expectation.
In the following we define an experimental cross section
\mbox{$\sigma^{\rm ano}_{f}= 0.943\times\sigma^{\rm pred,new}_{f}$~10$^{-43}$~cm$^2$/fission}.
With the new antineutrino spectra, we observe that for the data sample the minimum
$\chi^2$ is $\chi^2_{\rm min,data}=19.6$. The fraction of simple Monte-Carlo experiments with
$\chi^2_{\rm min}<\chi^2_{\rm min,data}$ is 25\%, showing that the distribution of
experimental ratios in $\overrightarrow{\text{R}}$ around the mean value is representative
given the correlations.

We will now discuss the possible explanations of this deviation from unity:
an erroneous prediction of the antineutrino flux
from the reactors, or a correlated artifact in the
experiments, or a real physical effect if both previous cases
are excluded.

Due to the importance of the antineutrino rate increase
we suggest that independent nuclear physics
groups should perform similar computations.
We also consider that new measurements of the electron spectra of
irradiated fissile isotopes would help clarifying the anomaly. All
cross sections of reactions used for the absolute and relative
normalizations of the ILL electron spectra have been checked and found
in agreement with the published values within error bars. A more
complete discussion on the evaluation of the normalization of reactor
antineutrino spectra based on the {\it ab initio} method will be
published later in~\cite{MurielAbInitio}.

Assuming the correctness of $\sigma^{\rm pred,new}_{f}$ the anomaly could
be explained by a common bias in all reactor neutrino experiments.
The measurements used one of two techniques, scintillator counters and integral detectors.
The Bugey-3 experiment~\cite{Bugey3} used 3 identical detection
modules, each of 600 liters, filled with $^6$Li-loaded liquid scintillator. Bugey-3
recorded 120,000 neutrino interactions.
The Bugey-4 experiment~\cite{Bugey4} used the Rovno91~\cite{Rovno91}
integral type detector, but increasing the antineutrino rate by a factor
of three. A similar detector was used for two integral measurements
Rovno88 1I and 2I~\cite{Rovno91}. In such detectors, based on $^3$He-filled counters,
the total number of antineutrino interactions is measured by
detecting only the neutrons from reaction Eq.~\ref{idb}.
The Goesgen experiment~\cite{Goesgen} used a detector nearly identical to the one
used for the ILL neutrino experiment~\cite{ILL81}, but with the additional feature
of position sensitivity. More than 10,000 neutrino events were
recorded at the three detector locations. The detector contained liquid scintillator
surrounded by $^3$He-filled wire chambers for neutron detection. Both the
positron and the neutron from reaction Eq.~\ref{idb} were detected.
Krasnoyarsk~\cite{Krasnoyarsk_87,Krasnoyarsk_94} used an integral
detector with a scintillation section. The Savannah River experiments
considered in this article used a scintillator counter~\cite{SRP96}.

Neutrons were tagged either by their capture in metal-loaded scintillator, or in
proportional counters, thus leading to two distinct systematics.
As far as the neutron detection efficiency calibration is concerned, we
note that different types of radioactive sources emitting MeV or sub-MeV
neutrons were used (Am-Be, $^{252}$Cf, Sb-Pu, Pu-Be).

It should be mentioned that the Krasnoyarsk, ILL, and SRP experiments
operated with nuclear fuel
such that the difference between the real antineutrino spectrum and
that of pure $^{235}$U was less than 1.5\%. They reported similar
deficits to those observed at other reactors operating with a mixed
fuel of $^{235}$U, $^{239}$Pu, $^{241}$Pu and $^{238}$U.
Hence the anomaly cannot be associated with a single fissile isotope.

All the elements discussed above argue against a trivial bias in the
experiments, but a detailed analysis of the most sensitive of them,
involving experts, would certainly improve the quantification of
the anomaly.

As discussed in Section~\ref{newsigmaf}, in a near future the reactor rate
anomaly significance might evolve due to the reevaluation of the cross
section prefactor $\kappa$ to 0.3961~$10^{-43}$~cm$^2$.  The averaged ratio
would shift to 0.938$\pm$0.023, leading to a deviation from unity at 99.2\%~C.L.

The third kind of possible explanation of the anomaly based on a real
physical effect will be detailed in Section~\ref{newphysics}.

\subsection{Shape information}
\label{anoshape} 

In this Section we re-analyze the Bugey-3 and ILL shape information,
based on the published data~\cite{Bugey3,ILL81}. We will use this
information for our combined analysis described in the next section.

\subsubsection{Bugey-3}

Based on the analysis of the shape of their energy spectra at
different source-detector distances~\cite{Bugey3,Goesgen},
the Goesgen and Bugey-3 measurements exclude oscillations such that
$0.06<\Delta m^2<1$~eV$^2$ for $\sin^2(2\theta)>0.05$.

\begin{figure}[!h]
\begin{center}
\includegraphics[scale=0.39]{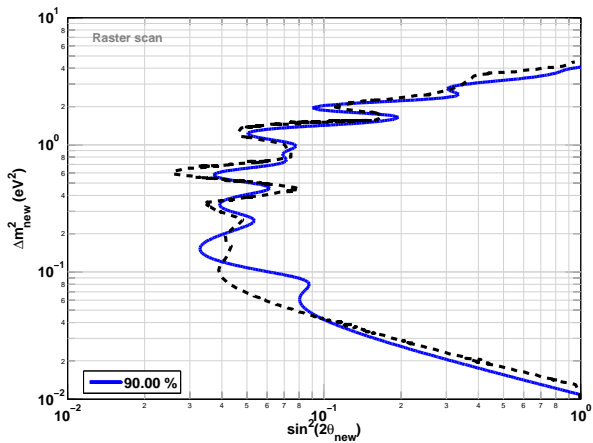}
\caption{\label{f:b3rep} 
90\%~C.L. exclusion domains obtained in the $\Delta
m^2$-$\sin^2(2\theta)$ plane from a raster scan of Bugey-3's data.
Our result (continuous line) is in good agreement with
the original result from~\cite{Bugey3} (dashed line), excluding
oscillations such that $0.06<\Delta m^2<1$~eV$^2$ for
$\sin^2(2\theta)>0.05$.}
\end{center}
\end{figure}

For further analysis we used Bugey-3's 40~m/15~m ratio data from~\cite{Bugey3}
as it provides the best limit. We followed the steps outlined in
\cite{Bugey3}, building the following $\chi^2$ function:
\begin{equation}
\label{eqb3}
\chi^2 = \sum_{i=1}^{N=25}
\left(\frac{(1+a)R_{\rm th}^i-R_{\rm obs}^i}{\sigma_i}\right)^2 +
 \left(\frac{a}{\sigma_a}\right)^2,
\end{equation}
where $R_{\rm obs}^i$ are our simulated data for Bugey from our tuned simulation
and $R_{\rm th}^i$ are our Monte-Carlo expectation in each bin. The $\sigma_i$ are
the errors reported by the Bugey-3 collaboration, and $a$ is a systematic parameter
accounting for the $\sigma_a=2\%$ uncertainty on the relative normalization at 40~m and 15~m.
Figure~16 of~\cite{Bugey3} shows the 90\%~C.L. exclusion contour from a {\it raster scan}
analysis with this estimator. As can be seen in Figure~\ref{f:b3rep}, we adequately
reproduce Bugey-3's results.

We already note here that in Section~\ref{newphysics}, when 
combining Bugey-3's shape information with other results,
we use the estimator from equation~\ref{eqb3}, but we perform a standard {\it global scan}
(i.e. the minimization is performed over the entire $\Delta m^2,\sin^2(2\theta)$ plane).
We only perform a raster scan here to show our agreement with the collaboration's
original analysis.

\subsubsection{ILL}

As already noted in Ref.~\cite{ILL95} the data from ILL
showed a spectral deformation compatible with an oscillation 
pattern in their measured over predicted events ratio. 
It should be mentioned that the parameters
best fitting the data reported by the authors of Ref.~\cite{ILL95} were
$\Delta m^2=2.2$~eV$^2$ and $\sin^2(2\theta)=0.3$. 

We reanalyzed the data of Ref.~\cite{ILL95} in order to include the
ILL shape-only information in our analysis of the reactor antineutrino
anomaly. We built the following shape-only estimator
\begin{equation}
\label{eqill}
\chi^2_{\mathrm{ILL,shape}}=\sum_{i=1}^{N=16} 
\left(\frac{(1+a)R_{\rm th}^i -
    R_{\rm obs}^i}{\sigma_i}\right)^2
\end{equation}
where $R_{\rm obs}^i$ are the measured ratios in each energy bin,
and $R_{\rm th}^i$ are our Monte-Carlo expectation in each bin.
$a$ is a free parameter in the fit, which renders this estimator
completely insensitive to any normalization information. It is therefore
only sensitive to the shape of the distribution. 

\begin{figure}[!h]
\begin{center}
\includegraphics[scale=0.39]{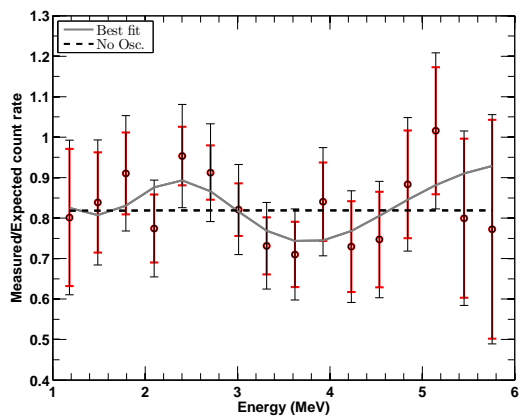}
\caption{\label{f:illshape} 
Ratio of measured to expected positron energy spectra of the ILL neutrino experiment (data points
extracted from~\cite{ILL95}). We show the best fit line with oscillations, along
with the no-oscillation line, from our shape-only fit.
The short error bars are statistical and the longer ones include the 11\% systematic
error described in the text.}
\end{center}
\end{figure}

$\sigma_i$ is the total error in each bin: we added in quadrature
the statistical error and a systematic error of 11\%.
It was difficult to extract the magnitude of this shape-only systematic error 
from published information. We combined our~$\chi^2_{\mathrm{ILL,shape}}$
estimator with the rate-only estimator, and verified that with this value
of the systematic, we could reproduce the ILL contours of both papers: 
when using $R_{\rm ILL,80}$ we reproduced
the contour in Figure~14 of Ref.~\cite{ILL81}, and when using~$R_{\rm ILL,95}$
we reproduced that of Ref.~\cite{ILL95} which excludes the no-oscillation
hypothesis at 2$\sigma$. 
A systematic error of 11\% is consistent with Figure~13 of Ref.~\cite{ILL81}.
It is also a conservative treatment of the shape-only information: with such
an error the data are compatible with the no-oscillation hypothesis at
1$\sigma$.
Figure~\ref{f:illshape} shows the data points from~\cite{ILL95}, along
with our best fit from the shape-only estimator 
($|\Delta m^2| \sim 2.3$~eV$^2$, $\sin^2(2\theta)
\sim 0.24$), and the no-oscillation line for comparison. 

\section{The fourth neutrino hypothesis}
\label{newphysics} 

In this section we discuss the compatibility of the reactor antineutrino
anomaly with the existence of a fourth non-standard neutrino,
corresponding in the flavor basis to the existence of a sterile
neutrino $\nu_s$ (see~\cite{pdg} and references therein).  
The motivation is the explanation of the antineutrino deficit by an
oscillation of electron neutrinos into a new neutrino state with a
large $\Delta m_{\rm new}^2$ value.

For simplicity we restrict our analysis to the 3+1 four-neutrino
scheme in which there is a group of three active neutrino masses separated
from an isolated neutrino mass, such that $|\Delta m_{\rm new}^2 |\gg 10^{-2}$
~eV$^2$. The latter would be responsible for very short baseline 
reactor neutrino oscillations.  
For energies above the inverse beta decay threshold and baselines
below 2 km, we adopt the approximated oscillation formula from 
Ref.~\cite{GouveaSterile}:
\begin{eqnarray}
P_{ee} & = & 1 -
\cos^4\theta_{\rm new}\sin^2(2\theta_{13})\sin^2\left(\frac{\Delta
    m_{31}^2L}{4E_{\nuebar}}\right) -  \nonumber \\
& & \sin^2(2\theta_{\rm new})\sin^2\left(\frac{\Delta
    m_{\rm new}^2L}{4E_{\nuebar}}\right).
\label{sterileproba}
\end{eqnarray}
In our analyses the well known solar driven
oscillation effects are negligible. The contribution of the
atmospheric driven oscillation is negligible at distances $<$ 100 m.
It is worth noting that we are not sensitive to any sterile neutrino 
mass hierarchy in the mass range considered. 

The ILL experiment may have seen a hint of oscillation in their
measured positron energy spectrum (see Section~\ref{ill}), 
but Bugey-3's results do not point to any significant spectral 
distortion more than 15~m away from the antineutrino
source. Hence, in a first approximation, 
hypothetical oscillations could be seen as 
an energy-independent suppression of the
$\nuebar$ rate by a factor of $\frac{1}{2}\sin^2(2\theta_{\rm new,R})$, thus leading to  
$\Delta m_{\rm new,R}^2 \gtrsim 1$ eV$^2$ and accounting for 
Bugey-3 and Goesgen shape analyses~\cite{Bugey3,Goesgen}.  
Considering the weighted averaged of all reactor experiments 
we get an estimate of the mixing angle,  $\sin^2(2\theta_{\rm new,R}) \sim
0.115$.

\begin{figure}
\begin{center}
	\includegraphics[scale=0.4]{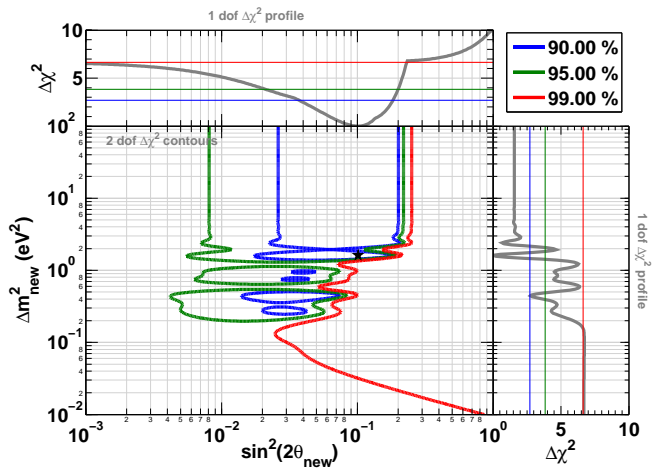}
	\caption{\label{f:sterile_r} Allowed  regions in the
          $\sin^2(2\theta_{\rm new})-\Delta m_{\rm new}^2$ plane obtained from
          the fit of the reactor neutrino data, without ILL-shape
          information, but with the stringent oscillation
          constraint of Bugey-3 based on the 40~m/15~m ratios
          to the 3+1 neutrino hypothesis, with $\sin^2(2\theta_{13})=0$. 
          The best-fit point is indicated by a star. 
        }
\end{center}
\end{figure}

Let us now fit the sterile neutrino hypothesis to the data (baselines
below 2~km) by minimizing the least-squares function 
\begin{equation}
\left( 
P_{ee}-\overrightarrow{\text{R}}
\right)^T 
W^{-1}
\left( 
P_{ee}-\overrightarrow{\text{R}}
\right), 
\end{equation}
assuming $\sin^2(2\theta_{13})=0$.

Figure~\ref{f:sterile_r} provides the results of the fit in the 
$\sin^2(2\theta_{\rm new})-\Delta m_{\rm new}^2$ plane, including the reactor
experiment rate information as well as the 40~m/15~m Bugey-3 spectral shape 
constraint presented in Figure~\ref{f:b3rep}. The latter leads to
stringent oscillation constraints for  $|\Delta m_{\rm new,R}^2 | \le 1$~eV$^2$,
since no spectral distortion was observed in the ratio 40~m/15~m. 
The fit to the data indicates that 
$|\Delta m_{\rm new,R}^2 | > 0.2$~eV$^2$ (95\%) and
$\sin^2(2\theta_{\rm new,R}) \sim 0.1$.
Note that if we include a non-zero value of $\sin^2(2\theta_{13})$, as large
as the 90\%~C.L. CHOOZ bound~\cite{Chooz}, the contours presented 
in Figure~\ref{f:sterile_r} are only marginally affected. 

\begin{figure*}
\begin{center}
	\includegraphics[scale=0.4]{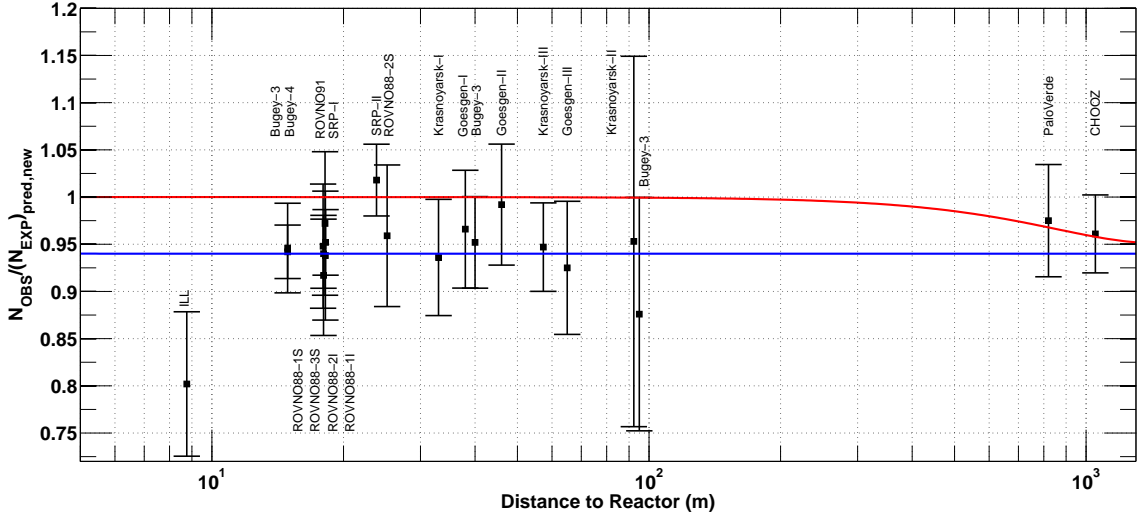}
	\caption{\label{f:anomalyfit} Illustration of the short
          baseline reactor antineutrino anomaly. The experimental results are
          compared to the prediction without oscillation, taking
          into account the new antineutrino spectra, the corrections
          of the neutron mean lifetime, and the off-equilibrium effects. 
          Published experimental errors and antineutrino spectra
          errors are added in quadrature. The mean averaged ratio
          including possible correlations is $0.943\pm 0.023$.
          The red line shows a possible 3~active neutrino mixing solution,
          with $\sin^2(2\theta_{13})=0.06$. 
          The blue line displays a solution including a new neutrino
          mass state, such as 
          $|\Delta m_{\rm new,R}^2 | \gg 1$~eV$^2$ and
          $\sin^2(2\theta_{\rm new,R})=0.12$ (for illustration purpose only).  
         }
\end{center}
\end{figure*}

Ignoring the reactor anomaly leads to an ambiguous interpretation of
the $\sim$1-2 km baseline experiment results constraining $\theta_{13}$.
As an example we compare the following two hypotheses: the 3~neutrino mixing scenario with 
$\sin^2(2\theta_{13})=0.06$ for instance and the 3+1
neutrino mixing scenario, taking $\sin^2(2\theta_{13})=0$, 
$\sin^2(2\theta_{\rm new,R})=0.115$ and $|\Delta m_{\rm new,R}^2 |>1$~eV$^2$
(best values of the combined fit shown on Figure~\ref{f:sterile_all}).  
Figure~\ref{f:anomalyfit} displays the data together with the two
hypotheses. The red line represents the three active neutrino
hypothesis, considering reactor neutrino rates until the CHOOZ baseline. 
The p-value of the 3~neutrino mixing hypothesis described above is 17\%
(26.9/21 degrees of freedom). 
The blue line illustrates the 3+1 neutrino mixing scenario
with a large $|\Delta m_{\rm new,R}^2 |=2.4$~eV$^2$, 
leading to a p-value of 55\% (19.4/21~degrees of freedom). 
We note a tension between the data and the three active neutrino
flavor hypothesis, again illustrating the anomaly.
Clear preference is given to the 3+1 neutrino hypothesis since $\Delta
\chi^2 \gtrsim 7$. A similar computation with the former reference
spectra would lead to a p-value of 34\% in both cases. 
This illustrates the impact of our
reanalysis with the new antineutrino spectra, updated neutron mean
lifetime, and off-equilibrium effects.

At this stage it is tempting to consider the previously noted
anomalies affecting other short baseline electron neutrino
experiments Gallex, Sage and MiniBooNE, reviewed in Ref.~\cite{GiuntiReview}.
Our goal is to quantify the compatibility with those anomalies.

\begin{figure*}[ht!]
\begin{center}
	\includegraphics[scale=0.39]{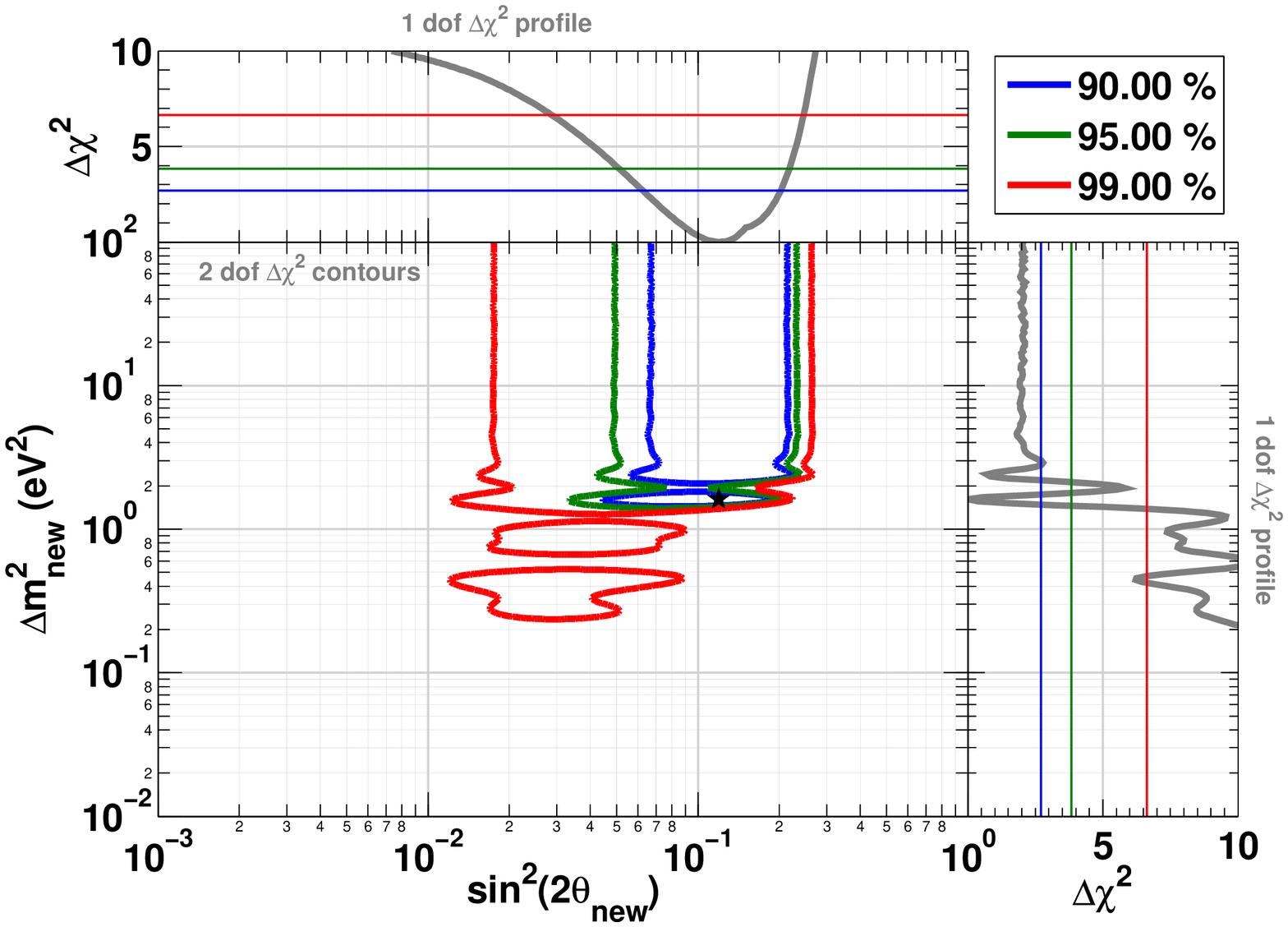}
	\includegraphics[scale=0.39]{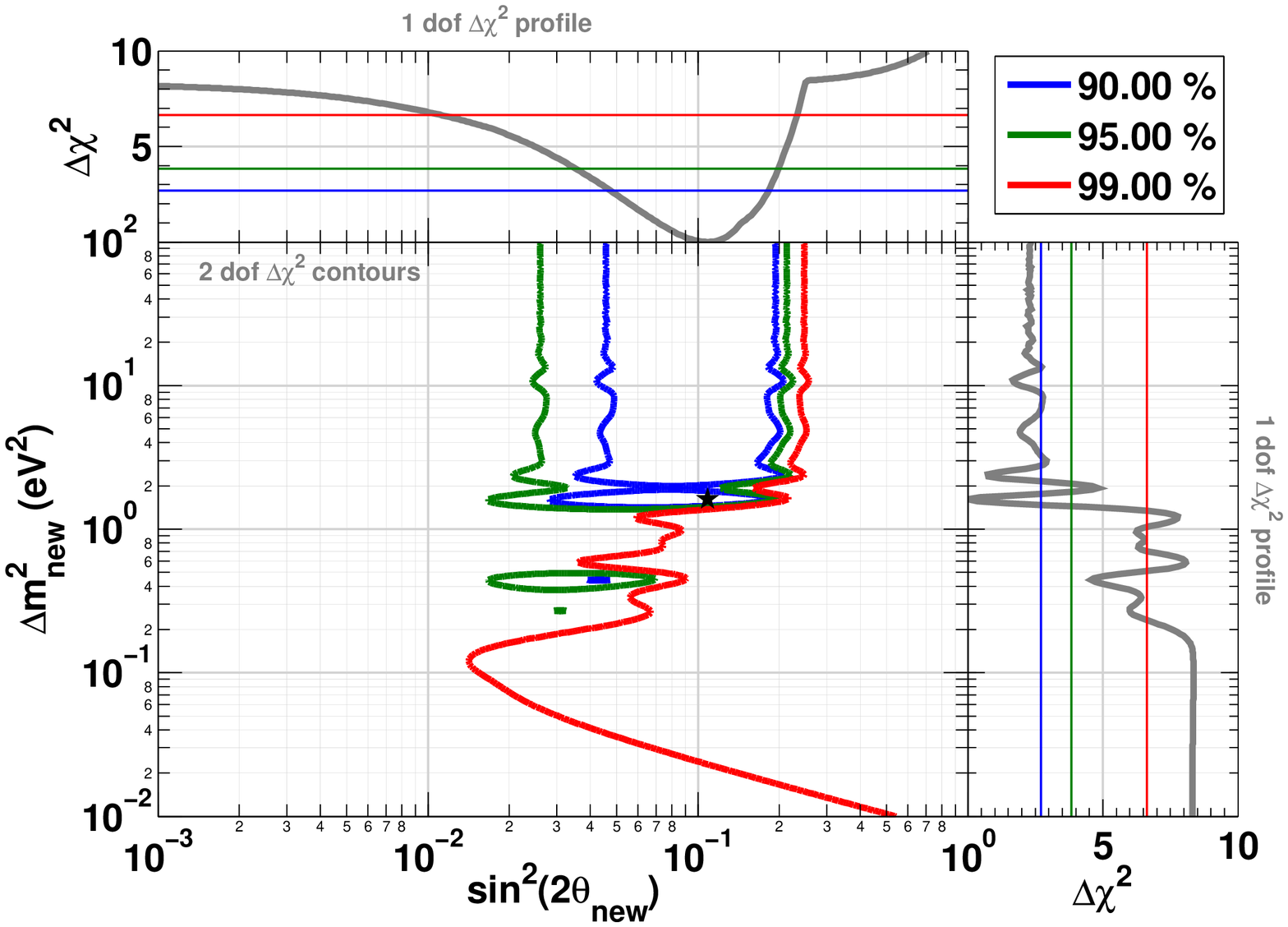}
	\caption{\label{f:sterile_rgrb} Allowed regions in the
          $\sin^2(2\theta_{\rm new})-\Delta m_{\rm new}^2$ plane obtained from
          the fit of the reactor neutrino data to the 3+1 neutrino
          hypothesis, with $\sin^2(2\theta_{13})=0$. 
          The left panel is the combination of the reactors and the
          gallium experiment calibration results with $^{51}$Cr 
          and $^{37}$Ar radioactive sources.
          The right panel is the combination of the reactors and our reanalysis of the
          MiniBooNE data following the method of
          Ref.~\cite{GiuntiReview}.
          In both cases the ILL energy spectrum information is not included.
        }
\end{center}
\end{figure*}

We first reanalyzed the Gallex and Sage calibration runs with $^{51}$Cr 
and $^{37}$Ar radioactive sources emitting $\sim 1$~MeV electron
neutrinos.~\cite{GallexSage}, following the methodology developed in
Ref.~\cite{GiuntiGallium, GiuntiReview}. However we decided to include possible 
correlations between these four measurements in this present work. 
Details are given in Appendix~\ref{reana}.
This has the effect of being slightly more conservative,
with the no-oscillation hypothesis disfavored at 97.7\%~C.L., 
instead of 98\%~C.L. in Ref.~\cite{GiuntiReview}.
Gallex and Sage observed an average deficit of \mbox{$R_G=0.86\pm 0.06\,(1\sigma)$}.
Considering the hypothesis of $\nu_e$ disappearance caused by
short baseline oscillations we used Eq.~(\ref{sterileproba}), 
neglecting the $\Delta m_{31}^2$ driven oscillations because of  
the very short baselines of order 1~meter.
Fitting the data leads to \mbox{$|\Delta m_{\rm new,G}^2| > 0.3$~eV$^2$} (95\%) and  
\mbox{$\sin^2(2\theta_{\rm new,G}) \sim  0.26$}.
Combining the reactor antineutrino anomaly with the gallium anomaly
gives a good fit to the data and disfavors the no-oscillation hypothesis 
at 99.7\%~C.L.
Allowed regions in the \mbox{$\sin^2(2\theta_{\rm new})-\Delta m_{\rm new}^2$} plane
are displayed in Figure~\ref{f:sterile_rgrb} (left).   
The associated best-fit parameters 
are \mbox{$|\Delta m_{\rm new,R\&G}^2| >1.5$~eV$^2$} (95\%) and  
\mbox{$\sin^2(2\theta_{\rm new,R\&G}) \sim 0.12$}.

\begin{table}[t]
\begin{center}
\medskip
\begin{tabular}{c|c|c|c}
\hline\hline
Experiment(s) &   $\sin^2(2\theta_{\rm new})$ &  $|\Delta m_{\rm new}^2 |$ (eV$^2$)   & C.L. (\%)\\
\hline
Reactors (no ILL-S,R$^*$) & 0.02-0.20 & $>0.40$   & 96.5\\
Gallium (G)               & $>0.06$   & $>0.13$  & 96.1\\
MiniBooNE (M)             & ---       & ---      & 72.4\\
ILL-S                     & ---       & ---      & 68.1\\
\hline
R$^*$ + G                 & 0.05-0.22 & $>1.45$  & 99.7\\
R$^*$ + M                 & 0.04-0.20 & $>1.45$  & 97.6\\
R$^*$ + ILL-S             & 0.02-0.21 & $>0.23$  & 95.3\\
\hline
All                       & 0.06-0.22 & $>1.5$   & 99.8\\
\hline\hline
\end{tabular}
\caption{\label{tab:summaryfit} Best fit parameter intervals or limits at 95\%~C.L. for
  $\sin^2(2\theta_{\rm new})$ and $|\Delta m_{\rm new}^2|$ parameters, and significance of the
  sterile neutrino oscillation hypothesis in \%, for different combinations of the reactor
  experimental rates only (R$^*$), the ILL-energy spectrum information
  (ILL-S), the gallium experiments (G), and MiniBooNE-$\nu$ (M)
  re-analysis of Ref.~\cite{GiuntiReview}. We quantify the difference
  between the $\sin^2(2\theta_{\rm new})$ constraints obtained from the
  reactor and gallium results. Following prescription of Ref.~\cite{Pgof},
  the parameter goodness-of-fit is 27.0\%, indicating reasonable agreement
  between the neutrino and antineutrino data sets (see Appendix~\ref{reana}).  
}
\end{center}
\end{table}

We then reanalyzed the MiniBooNE electron neutrino excess assuming the
very short baseline neutrino oscillation explanation of
Ref.~\cite{GiuntiReview}. 
Details of our reproduction of the latter analysis
are provided in Appendix~\ref{reana}.
The best fit values are \mbox{$|\Delta m_{\rm new,MB}^2|=1.9$~eV$^2$} and  
\mbox{$\sin^2(2\theta_{\rm new,MB}) \sim 0.2$}, but are not significant at 95\%~C.L.
The no-oscillation hypothesis is only disfavored at the
level of 72.4\%~C.L., less significant than the reactor
and gallium anomalies.
Combining the reactor antineutrino anomaly with our MiniBooNE
reanalysis leads to a good fit with the sterile neutrino hypothesis
and disfavors the absence of oscillations at 98.5\%~C.L.,
dominated by the reactor experiments data.
Allowed regions in the $\sin^2(2\theta_{\rm new})-\Delta m_{\rm new}^2$ plane
are displayed in Figure~\ref{f:sterile_rgrb} (right).
The associated best-fit parameters are
\mbox{$|\Delta m_{\rm new,R\&MB}^2 | > 0.4$~eV$^2$} (95\%) and
\mbox{$\sin^2(2\theta_{\rm new,R\&MB}) \sim 0.1$}.

\begin{figure}
\begin{center}
	\includegraphics[scale=0.39]{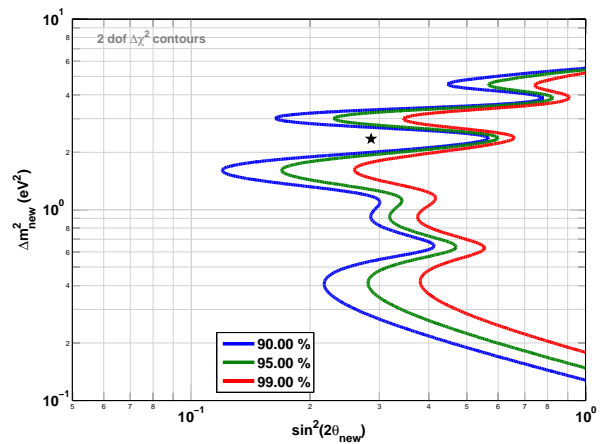}
	\caption{\label{f:sterile_illshape} Allowed regions in the
          $\sin^2(2\theta_{\rm new})-\Delta m_{\rm new}^2$ plane obtained from
          a fit of the ILL energy spectrum shape only.
          The best fit value reported by the authors of
          Ref.~\cite{ILL95} is very close to our best fit,
          at $|\Delta m_{\rm new}^2 |\sim$2 eV$^2$, but it is worth noting its
          poor statistical significance, compatible with the absence
          of oscillations at the 1$\sigma$ level.
          The best-fit point is indicated by a star.
        }
\end{center}
\end{figure}

\begin{figure*}
\begin{center}
	\includegraphics[scale=0.6]{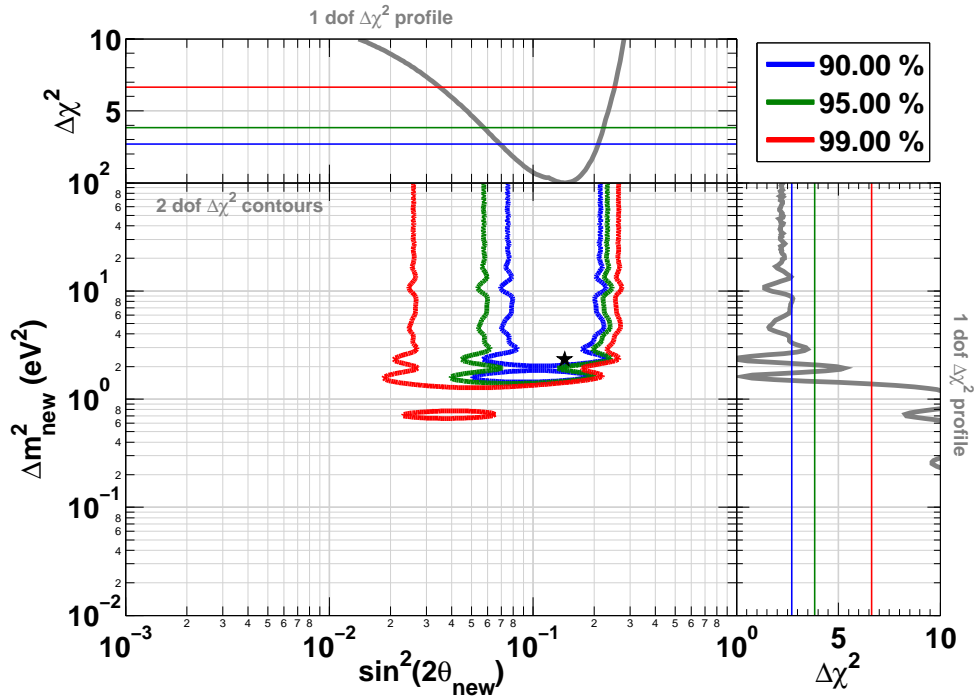}
	\caption{\label{f:sterile_all} Allowed regions in the
          $\sin^2(2\theta_{\rm new})-\Delta m_{\rm new}^2$ plane from the
          combination of reactor neutrino experiments, Gallex and Sage
          calibration sources experiments, MiniBooNE reanalysis of
          Ref.~\cite{GiuntiReview}, and the ILL-energy spectrum distortion.
          The data are well fitted by the 3+1 neutrino hypothesis,
          while the no-oscillation hypothesis is disfavored at 99.8\%~C.L.
          The marginal $\Delta \chi^2$ profiles for $|\Delta m_{\rm new}^2|$ and
          $\sin^2(2\theta_{\rm new})$ (1~dof) lead to the constraints,
          $|\Delta m_{\rm new}^2 | > 1.5$~eV$^2$ (95\%~C.L.) and
          $\sin^2(2\theta_{\rm new})=0.14 \pm 0.08$ (95\%~C.L.).
        }
\end{center}
\end{figure*}

Our ILL re-analysis, including only the energy spectrum shape,
leads to the allowed regions in the $\sin^2(2\theta_{\rm new})-\Delta
m_{\rm new}^2$ plane presented in Figure~\ref{f:sterile_illshape}.
We notice a hint of neutrino oscillations such that
\mbox{$|\Delta m_{\rm new,ILL-shape}^2| > 1$~eV$^2$} and
\mbox{$\sin^2(2\theta_{\rm new,ILL-shape}) \sim 0.2$},
in agreement with our fourth neutrino hypothesis, but still
compatible with the absence of oscillations at the 1$\sigma$ level.
Figure~\ref{f:illshape} is our reproduction of the
illustration~3 of Ref.~\cite{ILL81}; we superimposed
the oscillation pattern that would be induced by neutrino oscillations
at our best fit (combined analysis).
The ILL positron spectrum is thus in agreement with the oscillation parameters
found independently in our re-analyses, mainly based on rate information.
Because of the differences in the systematic effects in the rate and shape
analyses, this coincidence is in favor of a true physical effect rather than
an experimental anomaly.
As a cross check we performed a Monte-Carlo simulation of the ILL and Bugey-3
experiments, including the finite spatial extension of the nuclear reactors and
the ILL and Bugey-3 detectors. We found that the small dimensions of
the ILL nuclear core lead to small corrections of the oscillation
pattern imprinted on the positron spectrum.
However the large extension of the Bugey nuclear
core is sufficient to wash out most of the oscillation pattern at 15~m.
This explains the absence of shape distortion in the Bugey-3 experiment.

Table~\ref{tab:summaryfit} summarizes all the results of our fits of
reactor, gallium, and MiniBooNE-$\nu$ data to the sterile neutrino
oscillation hypothesis. We observe that all the data sets taken separately
are very consistent with one another, pointing to very similar oscillation parameters.
We thus performed a global fit to all available data.

The no-oscillation hypothesis is disfavored at 99.8\%~C.L.
The significance is dominated by the gallium and reactor
data. Allowed regions in the $\sin^2(2\theta_{\rm new})-\Delta m_{\rm new}^2$ plane
are displayed in Figure~\ref{f:sterile_all}, together with the
marginal $\Delta\chi^2$ profiles for $|\Delta m_{\rm new}^2 |$ and
$\sin^2(2\theta_{\rm new})$. The combined fit leads to the following constraints on
oscillation parameters: $|\Delta m_{\rm new}^2|>1.5$~eV$^2$ (95\%~C.L.)
and $\sin^2(2\theta_{\rm new})=0.14\pm0.08$ (95\%~C.L.).
An embryo of possible consequences of this result will
be discussed in Section~\ref{conclusion}.

\section{Revision of the constraints on $\theta_{13}$}
\label{theta13}

In this section we discuss the impact of our revision of the results
of short baseline ($<$~100 m) reactor experiments on the
constraints on the $\theta_{13}$ mixing angle from CHOOZ~\cite{Chooz},
KamLAND~\cite{Kamland}, and their combination.
The results depend on the choice of the cross section per
fission used to normalize the predicted antineutrino rate, as well as
on the neutrino oscillation scheme used in the analysis, i.e. whether to
include a non-standard neutrino or not.
Assuming the correctness of the neutrino experiments (detected rate),
we show that constraints on $\theta_{13}$ could be derived by using
$\sigma^{\rm pred,ano}_{f}$ and the three-active neutrino scenario.

\subsection{CHOOZ}
\label{chooz}

\begin{figure}[!h]
\begin{center}
\includegraphics[scale=0.39]{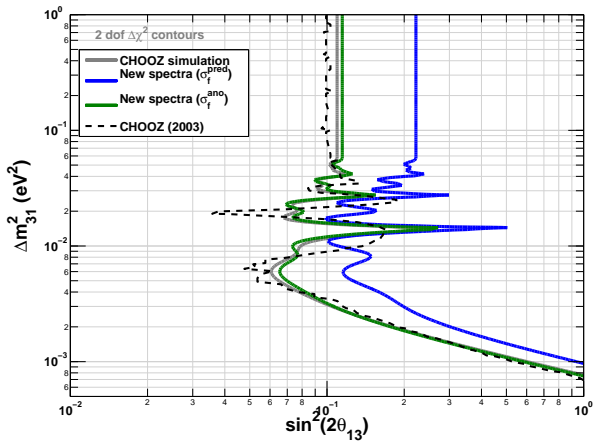}
\caption{\label{fig:chooz} 
Our simulation (gray line) roughly reproduces the published CHOOZ exclusion
limit~\cite{Chooz} represented by the black dashed line.
We obtain the blue contour by normalizing the expected no-oscillation antineutrino
rate to $\sigma^{\rm pred,new}_{f}$ and analyzing the data using the three
active neutrino oscillations scheme.
We then normalize the expected antineutrino event rate to $\sigma^{\rm ano}_{f}$
to account for possible neutrino oscillations at very short baselines,
leading to the green contour. This provides our new constraint on
$\sin^2 (2\theta_{13})$, accounting for the reactor antineutrino anomaly.
}
\end{center}
\end{figure}

We first reproduced CHOOZ's background-subtracted results and Monte-Carlo
using Ref.~\cite{Chooz}. We used the converted neutrino spectra from
ILL electron data~\cite{SchreckU5_85, Vogel81, Kopeikin88}, using the
parameterization from Ref.~\cite{Huber:2004xh},
and the \mbox{V-A}~cross section from~\cite{FAY85, VogelXsec}, normalized to the
Bugey-4 value as was done in Ref.~\cite{Chooz}.
Long-lived fission product contributions were also taken into account
(at 300~days of irradiation).
As in Ref.~\cite{Chooz}, we accounted for a $\sigma_\alpha=2.7\%$ systematic error
on the overall normalization and a~1.1\% error on the energy scale,
along with a bin-to-bin uncorrelated uncertainty
describing the uncertainty of the neutrino spectrum.
Using the original ILL neutrino converted spectra, with the
$\sigma^{\rm Bugey}_{f}$ cross-section we could roughly reproduce
the exclusion limit from Ref.~\cite{Chooz} (black dashed
line on Figure~\ref{fig:chooz}) to acceptable precision (gray line on
Figure~\ref{fig:chooz}) in the relevant $\Delta m_{31}^2$ range,
although we could not fully imitate
every step in the published analysis. In particular we did not
use the same statistical treatment, using only a global scan
while the CHOOZ collaboration used the Feldman-Cousins unified approach
to extract the confidence interval.

Let us now revisit the CHOOZ results using the
new antineutrino spectra in the three active neutrino framework.
We perform the same analysis as before,
but computing the expected number of events based
on $\sigma^{\rm pred,new}_{f}$ and its error. $\sigma_\alpha$
was thus increased to~3.3\% to account for this effect.
This leads to a new exclusion limit, $\sin^2 (2\theta_{13})<0.22$
(1~dof) at~90\%~C.L. for \mbox{$\Delta m^2_{31}=2.4 \, 10^{-3}$~eV$^2$}.
The corresponding exclusion contour is displayed as the blue line on
Figure~\ref{fig:chooz}.
In this approach we make use of $\sigma^{\rm pred,new}_{f}$
and attribute the slight deficit of $\nuebar$ at CHOOZ
to a $\theta_{13}$ driven oscillation, such that
$\sin^2 (2\theta_{13})=0.11^{+0.07}_{-0.08}$.
However this result only holds within the
three active neutrino mixing framework,
under the hypothesis of no-oscillation at very short baselines.
But in Section~\ref{anomaly} we pointed out an anomaly affecting
the short baseline reactor neutrino experiment results.
In the scenario discussed above with a new neutrino ($\Delta{m}^2_{\rm new}$,
$\theta_{\rm new}$) we generate oscillations driven by a large
$|\Delta m_{\rm new}^2| \gg \Delta m_{31}^2$. In this case CHOOZ measured
a combination of the oscillation driven by $(\Delta
m_{\rm new}^2,\theta_{\rm new})$ and $(\Delta m_{31}^2,\theta_{13})$.
A comprehensive 3+1 neutrino analysis would thus be mandatory to
constrain $\theta_{13}$ using the new normalization based on
$\sigma^{\rm pred,new}_{f}$, but we will see next that another approach
may be used.

As a matter of fact the CHOOZ experiment used
$\sigma^{\rm Bugey}_{f}$ instead of $\sigma^{\rm pred,new}_{f}$. The
collaboration justified making use of $\sigma^{\rm Bugey}_{f}$ because of
the good agreement with $\sigma^{\rm pred,old}_{f}$. With our
revision of $\sigma^{\rm pred}_{f}$ this justification does not hold
anymore in the 3 neutrino oscillation framework.
But it turns out that if we assume non-standard neutrino oscillations
in the large $|\Delta m_{\rm new}^2 |\gg 0.1$ eV$^2$  regime (no spectral distortion),
the normalization of CHOOZ using $\sigma^{\rm Bugey}_{f} \sim
\sigma^{\rm ano}_{f}$ leads to an estimator of $\sin^2 (2\theta_{13})$
which eliminates a possible degeneracy with $\sin^2 (2\theta_{\rm new})$.
CHOOZ's strategy was indeed to absorb
possible errors in rate predictions, but this methodology holds as well
for constraining $\theta_{13}$ on top of an additional, short baseline,
averaged oscillation.

Thus a pragmatic approach for constraining $\theta_{13}$ from CHOOZ
data is the use of the weighted average of the measured cross sections of all experiments
$<100$~m, $\sigma^{\rm ano}_{f}$.
This leads to a new exclusion limit, $\sin^2(2\theta_{13})<0.10$ (1~dof)
at~90\%~C.L. for $\Delta m^2_{31}=2.4 \, 10^{-3}$~eV$^2$,
slightly lower than CHOOZ's published value~\cite{Chooz}.

\subsection{KamLAND}
\label{kamland}

In this section we reevaluate the KamLAND constraint on~$\theta_{13}$
in light of the reactor antineutrino anomaly.

Unlike in the original CHOOZ analysis from~\cite{Chooz},
the KamLAND $\nuebar$ energy spectrum is
calculated using $\sigma^{\rm pred,old}_{f}$ based on Ref.~\cite{SchreckU5_85,Vogel81}.
But the spectral uncertainty is evaluated from Ref.~\cite{Bugey4},
including off-equilibrium effects. In this case
the propagation of the Bugey-4 error on $\sigma^{\rm pred}_{f}$ only has
a marginal impact since it is not among the dominant systematics.

We first reproduced KamLAND's results and Monte-Carlo to a very good
accuracy exploiting Ref.~\cite{Kamland} as well as publicly available
information~\cite{IAEAdb}. We used the parameterization of
Ref.~\cite{Huber:2004xh} for $S_{\rm tot}$, and included
off-equilibrium corrections according to Ref.~\cite{LM}. 
Special care was taken to include all published backgrounds,
known antineutrino sources, especially Korean power reactors, and geoneutrinos.
We tuned our simulation to reproduce KamLAND's Monte-Carlo with and
without neutrino oscillations (adjusting each effective power of the
Japanese power plants).
In both cases our simulation agrees with that of KamLAND to better than~1\% in the
1.5-6.5~MeV range.

Confidence levels in the $(\tan^2\theta_{12},\sin^2\theta_{13})$ plane
were obtained by minimizing the generic $\chi^2$ function:
\begin{equation}
\chi^2 = \sum_{i} \left(\frac{Y_i - N_i(\alpha,\beta)}{Y_i}\right)^2 + 
\beta^{T}W^{-1}\beta
\end{equation}
where the $Y_i$ are our simulated data tuned to KamLAND's Monte-Carlo~\cite{Kamland}.
The free parameters are $\alpha$ ($\theta_{12}$, $\theta_{13}$,
$\Delta{m}^2_{21}$ and geo-$\nu$ rate), and the nuisance
parameters, $\beta$, which are the systematics quoted in Ref.~\cite{Kamland}.
The $N_i(\alpha,\beta)$ are our simulation model for all the free
parameters. The $W^{-1}$ matrix contains all the systematic
uncertainties quoted in Ref.~\cite{Huber:2004xh}.

\begin{figure}[!h]
\begin{center}
\includegraphics[scale=0.39]{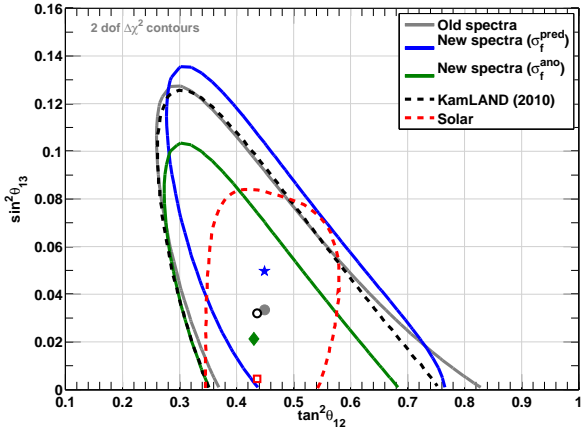}
\caption{\label{fig:kamland} 
Our simulation (gray line, gray disk) reproduces the published 95\%~C.L.
KamLAND inclusion contour~\cite{Kamland} represented by the black
dashed line (black circle).
We obtain the blue contour (star) by normalizing the expected no-oscillation antineutrino
rate to $\sigma^{\rm pred,new}_{f}$ and analyzing the data using the three
active neutrino oscillations scheme.
We then normalize the expected antineutrino event rate using $\sigma^{\rm ano}_{f}$
to account for possible neutrino oscillations at very short baselines,
leading to the green contour (diamond). This provides our new constraints in
the $\tan^2\theta_{12}$-$\sin^2\theta_{13}$ plane,
accounting for the reactor antineutrino anomaly. We note that the
slight tension between the $\sin^2\theta_{13}$ best-fit values of
solar (red square) and reactor data is reduced in our scenario.
}
\end{center}
\end{figure}

Figure~\ref{fig:kamland} demonstrates that we
could reproduce the best fit and the confidence contours of
KamLAND~\cite{Kamland} with good accuracy.
With the original reactor neutrino spectra $S_{\rm tot}$,  we obtain a
good agreement with KamLAND's published results~\cite{Kamland}.
Changing the reference spectra $S_{\rm tot}$ according to Ref.~\cite{LM},
the best fit values and uncertainties on $\tan^2\theta_{12}$ and
$\Delta{m}^2_{21}$ are unaffected in a three active neutrino,
oscillations framework, but the $\sin^2(2\theta_{13})$ central
value is shifted upwards to
$\sin^2(2\theta_{13})=0.2_{-0.13}^{+0.14}$, consistent with zero at $1.5\sigma$.
The new 90\%~C.L. limit on $\sin^2(2\theta_{13})$ would
therefore increase to $0.41$ (1~dof), marginalizing over the other
fit parameters.

However, as in the case of CHOOZ discussed above, these results do
not take into account the reactor antineutrino anomaly at short baselines.
In the 3+1 neutrino oscillation framework, the whole effect induced by the
normalization shift is absorbed by the new
oscillation at very short baselines, driven by $\Delta{m}^2_{\rm new}$.
Normalizing KamLAND's data to $\sigma^{\rm ano}_{f}$ leads to an estimator
revising the constraint on $\theta_{13}$ using the three-active
neutrino oscillation framework, such that
$\sin^2(2\theta_{13})<0.31$ (1~dof) at
90\%~C.L. for $\Delta m^2_{31}=2.4 \, 10^{-3}$~eV$^2$.

\subsection{Combining CHOOZ and KamLAND}
\label{combined}

\begin{figure*}
\begin{center}
	\includegraphics[scale=0.39]{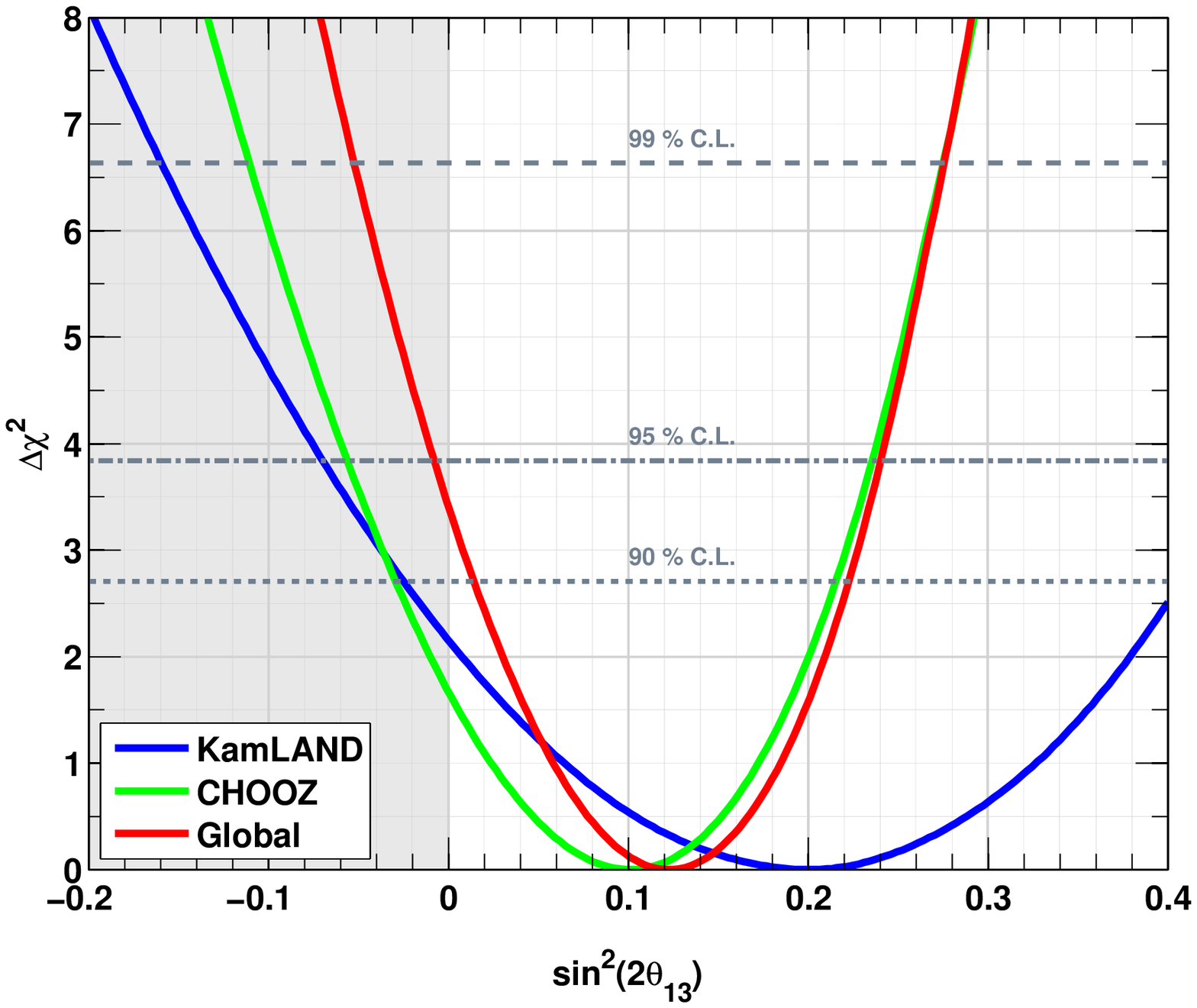}
	\includegraphics[scale=0.39]{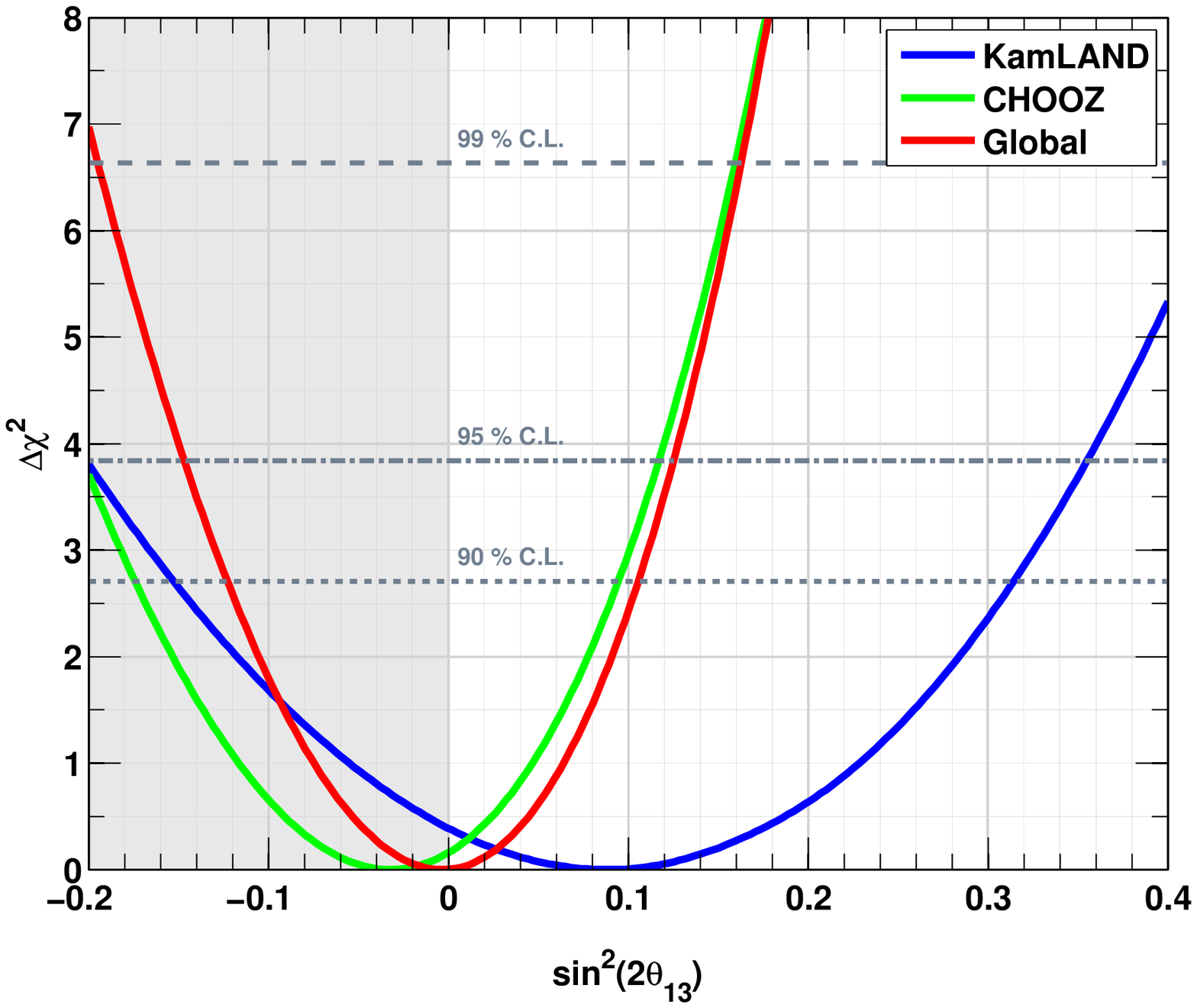}
	\caption{\label{DeltaChi2_s2_t13} Dependence of $\Delta\chi^2$ on
  $\sin^2(2\theta_{13})$ for CHOOZ (green), KamLAND (blue), and their
  combination (red). 
  Comparing the data with the excess in expected event rate predicted 
  using $\sigma^{\rm pred,new}_{f}$ leads to an increase of the best-fit
  value of $\sin^2(2\theta_{13})$, in the three active neutrino oscillations
  framework (left). 
  The normalization of the expected event rate with $\sigma^{\rm ano}_{f}$ 
  allows to absorb the reactor antineutrino anomaly observed at very short
  baseline, leading to our revised constraints on $\sin^2(2\theta_{13})$.
        }
\end{center}
\end{figure*}

In sections \ref{chooz} and \ref{kamland} we separately revisited the 
CHOOZ and KamLAND results.
Here, we first voluntarily ignore the reactor antineutrino anomaly at short
baselines. We thus normalize both CHOOZ and KamLAND with $\sigma^{\rm pred,new}_{f}$. 
Combining these two results leads to a best-fit value of $\sin^2(2\theta_{13}) =
0.13^{+0.06}_{-0.06}$, barely consistent with a null value of~$\theta_{13}$.
The left panel of Figure~\ref{DeltaChi2_s2_t13} shows the $\Delta\chi^2$ profiles
for CHOOZ (green), KamLAND (blue), and for their combination (red), as a
function of $\sin^2(2\theta_{13})$, marginalizing over the other parameters.
In this scenario the revised limit is $\sin^2(2\theta_{13})<0.23$ at~90\%~C.L. (1~dof),
for the estimation of~$\theta_{13}$ only, 
for $\Delta m^2_{31}=2.4 \, 10^{-3}$~eV$^2$.
These results are consistent with analyses of various data sets
indicating a `hint' for a non-zero
$\theta_{13}$~\cite{LisiHint08,theta13review}. 
This is explained by the increase of the predicted
event rate at CHOOZ and KamLAND, which is attributed to $\theta_{13}$
only, in the three neutrino oscillations framework.

But the previous estimation does not take into account the reactor
antineutrino anomaly. 
Let us now consider that both CHOOZ and KamLAND are normalized 
with $\sigma^{\rm ano}_{f}$,  in order to constrain $\theta_{13}$ using a
three active neutrino oscillation framework.
The right panel of Figure~\ref{DeltaChi2_s2_t13} shows the $\Delta\chi^2$ profiles
for CHOOZ (green), KamLAND (blue), and for their combination (red), 
computed as before. 
In this scenario the best fit is at $\sin^2(2\theta_{13}) =0.0^{+0.07}_{-0.07}$. 
This lower best fit value is due to the fact that we predict fewer
antineutrinos at CHOOZ and KamLAND, some of them having already
oscillated into non-standard neutrinos.  
This leads to our revised constraint, $\sin^2(2\theta_{13})<0.11$ at
90\%~C.L. (1~dof) for $\Delta m^2_{31}=2.4 \, 10^{-3}$~eV$^2$.
Results are summarized in Table~\ref{tab:global}.

We conclude that the hint for a non zero value of~$\theta_{13}$,
first presented in Ref.~\cite{LisiHint08}, vanishes if one normalizes 
CHOOZ's and KamLAND's predicted antineutrino rates to $\sigma^{\rm ano}_{f}$. 
Furthermore we notice that the slight tension affecting the $\sin^2\theta_{13}$
best-fit values of solar and reactor
data~\cite{theta13review} is reduced in our scenario involving
 a non-standard neutrino such that $\Delta m^2_{\rm new} \gg \Delta
 m^2_{31}$.

A full analysis of the data based on non-standard neutrino oscillation
models would lead to more accurate results, beyond the scope of this article.

\begin{table}[t]
\begin{center}
\medskip
\begin{tabular}{c|c|c|c|c}
\hline\hline
                   &  \multicolumn{2}{|c|}{$\sigma^{\rm pred,new}_{f}$}  &   \multicolumn{2}{c}{$\sigma^{\rm ano}_{f}$}   \\
\hline
 Re-analysis & Best fit$^{+1\sigma}_{-1\sigma}$ & 90\%~C.L. & Best fit$^{+1\sigma}_{-1\sigma}$ & 90\%~C.L. \\
%&& (90\%~C.L.)\\
&&&&\\[-2mm]
\hline
&&&&\\[-2mm]
KamLAND  & $0.20^{+0.13}_{-0.14}$ & $<0.41$ & $0.09^{+0.14}_{-0.14}$  & $<0.31$\\[1.5mm]
CHOOZ    & $0.11^{+0.07}_{-0.08}$ & $<0.22$ & $-0.02^{+0.08}_{-0.08}$ & $<0.10$\\[1.5mm]
Combined & $0.13^{+0.06}_{-0.06}$ & $<0.23$ & $0.00^{+0.07}_{-0.07}$ & $<0.11$\\[1.5mm]
\hline\hline
\end{tabular}
\caption{\label{tab:global} Best-fit values for $\sin^2(2\theta_{13})$
  and $1\sigma$ errors, considering two possible normalizations with
$\sigma^{\rm pred,new}_{f}$ or $\sigma^{\rm ano}_{f}$. Limits are given for
  a 1~dof parameter estimation.}
\end{center}
\end{table}

\subsection{Daya Bay, Double Chooz, and RENO}
\label{dcdbre}

Three new reactor neutrino oscillation experiments, Daya
Bay~\cite{Dayabay},  Double Chooz~\cite{Doublechooz}, 
and RENO~\cite{Reno} will soon be coming online to significantly 
improve our knowledge of reactor antineutrino rates at various 
baselines.

Among these experiments Double Chooz is the only one to operate a
first phase without a near detector. 
The evidence for a reactor antineutrino anomaly triggers the
discussion of the choice of the cross section per fission to be used
for constraining $\theta_{13}$, $\sigma^{\rm pred,new}_{f}$ or
$\sigma^{\rm ano}_{f}$, as well as its associated uncertainty.

Straightforwardly using $\sigma^{\rm pred,new}_{f}$ in the three neutrino
oscillation framework to interpret a hypothetical antineutrino deficit
at Double Chooz could lead to an overestimation of $\theta_{13}$, which
could fake a discovery. But if the sterile neutrino oscillation
hypothesis discussed in this paper is proven, the use of  
$\sigma^{\rm pred,new}_{f}$ would be possible in active+sterile 
neutrino oscillation frameworks. In that case the~$2.7\%$ error budget
of the antineutrino spectra should be used.
Assuming that the average experimental results of reactor neutrino are
correct, using $\sigma^{\rm ano}_{f}$ has the advantage to absorb 
either a antineutrino flux miscalculation, or a physical neutrino
deficit at very short baselines, leading to a conservative constraint
on $\theta_{13}$. The error budget could then be taken as the
weighted standard deviation of the short baseline experiments,~1.0\%.

This choice is however not relevant for experiments running with a
multi-detector configuration, since they absorb part of the uncertainty of the
reactor antineutrino fluxes, depending on the setup. 
Even in the hypothetical case of antineutrino oscillation at very
short baselines the sensitivities of Daya
Bay, Double Chooz, and RENO should be marginally affected 
because of the large allowed values of $\Delta m_{\rm new}^2$. However
this statement should be checked based on detailed simulations.

In the non-standard neutrino hypothesis the discovery of a 
shape distortion in the positron energy spectrum 
of the far detectors may be determinant in disentangling~$\theta_{13}$
from~$\theta_{\rm new}$.

\section{Testing the anomaly}
\label{testanomaly}

\subsection{At reactors}
\label{testreactors} 
The presence of sterile neutrinos would leave its imprint on the
signal at both the near and far detectors of forthcoming reactor 
neutrino experiments. 
Taking results from multi-detectors and
allowing for the possibility of sterile mixing angles, one can probe
both $\theta_{13}$ and the sterile mixing angle $\theta_{\rm new}$ at Daya
Bay, Double Chooz, and RENO~\cite{theta1314}. Positron spectrum energy
distortions should be deeply investigated at near detectors, as quoted in~\cite{GouveaSterile}. 
In any case, measurements of the expected over predicted event rates
at the near detectors will allow the probing of the reactor
antineutrino anomaly, providing high statistics and high precision
measurements at a few hundreds meters from the antineutrino sources. 
The antineutrino anomaly would be best tested by
performing blind analyses for all near detector data.

Further measurements at very short baselines below 100~meters would be
useful to confirm the MeV electron antineutrino deficit.
Currently no fundamental research program is underway to search for 
new oscillation physics at reactors. 
However there is a worldwide program at short baselines for the
purpose of nuclear non-proliferation, using antineutrinos as new IAEA safeguards
tools~\cite{aiea2008report}. In this context the Nucifer experiment,
located 7~meters away from the Osiris research reactor core in
Saclay, will start its operation in 2011~\cite{Nucifer}. Nucifer will
thus have the possibility to test the anomaly. A rate-only analysis 
with a precision of a few percent may not be enough to provide a 
decisive improvement of the understanding of the anomaly. 
But a shape analysis may provide enough information, depending of the
energy resolution. 

We note here that the antineutrino non-proliferation program will not
be affected by the antineutrino anomaly since relative antineutrino
rates with respect to known thermal powers could be used to calibrate the
experiments.
 
\subsection{MegaCurie radioactive sources}
\label{testarcal} 
As mentioned above, radiochemical gallium experiments (Gallex and
Sage) tested their experimental procedure by exposing their gallium
target to MegaCurie neutrino sources using $^{51}$Cr or $^{37}$Ar~\cite{GallexSage}. 
The production and handling of such devices is thus well under
control. 
Ref.~\cite{GavrinSterile} proposed to use a $^{51}$Cr source inside
two concentric gallium tanks, whereas Ref.~\cite{IanniSterile} proposed
to use a $^{51}$Cr or $^{90}$Sr source next to the Borexino detector.
In liquid scintillating detectors $^{51}$Cr or $^{37}$Ar $\nu_e$'s
are detected through neutrino-electron elastic scattering while 
$^{90}$Sr $\nuebar$'s are detected through inverse beta decay. 
With the $\Delta m_{\rm new}^2$ values best fitting the sterile
neutrino hypothesis, the deployment of a radioactive source at the
center of an ultra-low background neutrino detector, such as Borexino,
KamLAND, and SNO+, would allow both the testing of the $\nu_e$
deficit and the search for an oscillation pattern as function of the detector
radius. These neutrino sources emit quasi-mono-energetic 
neutrino lines of sub-MeV energies leading to a clear oscillation
pattern at the range of a meter. In addition a $^{37}$Ar source
emits only $\gamma$-rays through second order processes and is
therefore easy to handle after its irradiation inside a breeder nuclear reactor.
As an example, if a source of 1~MCi were inserted in the middle of a
large detector like SNO+, it would provide a few hundred thousand
interactions in the detector, of which a few ten thousands 
would deposit more than 500~keV. A $<15$~cm accurate vertex
reconstruction could allow to draw a simple and clear figure of the
number of neutrino interactions as a function of radius, directly 
testing the sterile neutrino oscillation pattern for 
$\Delta m_{\rm new}^2<5$~eV$^2$.

\section{Conclusions and Outlook}
\label{conclusion}

The impact of the new spectra of reactor antineutrino is extensively 
studied in this article. The increase of the expected antineutrino 
rate by about~3.5\% combined with revised values of the antineutrino
cross section significantly decreased the normalized ratio of observed to expected event
rates in all previous reactor experiments performed over the last 30~years
at distances below 100~m~\cite{ILL81,Goesgen,Rovno88, Rovno91,Krasnoyarsk_87,Krasnoyarsk_94,Bugey3,Bugey4,Paloverde,Chooz}. 
The average ratio is $0.943\pm 0.023$, leading to the reactor
antineutrino anomaly. This deficit
could still be due to some unknown in the reactor physics, but we
also analyze these revised results in terms of a suppression of the
$\nuebar$ rate at short distance as could be expected from a sterile neutrino,
beyond the standard model, with a large
$|\Delta m_{\rm new}^2| \gg |\Delta m_{31}^2|$. 
We note that hints of such results were already present at the ILL
neutrino experiment in~1981~\cite{ILL95}.
We also considered that other neutrino experiments,
MiniBooNE~\cite{miniboone} and the gallium neutrino sources 
experiments~\cite{GallexSage}, observe $\nu_e$ 
deficits at a similar $L/E$. These anomalies were comprehensively
studied in Ref.~\cite{GiuntiReview}. 
It is important to note that these anomalies exist and are
comparable in both the neutrino and the antineutrino sectors.
Furthermore it turns out that each experiment
fitted separately leads to similar values of $\sin^2(2\theta_{\rm new})$
and similar lower bounds for $|\Delta m_{\rm new}^2|$, but without a 
strong significance. Hence, we combined in a global fit these results
 with our re-evaluation of the reactor experimental results, 
taking into account the existing correlations. 
This leads to a solution for a new neutrino oscillation, 
such that $|\Delta m_{\rm new}^2|>1.5$~eV$^2$ (95\%~C.L.) and 
$\sin^2(2\theta_{\rm new})=0.14\pm 0.08$ (95\%~C.L.), disfavoring
the no-oscillation case at 99.8\%~C.L.
This hypothesis should be checked against systematical effects, either 
in the prediction of the reactor antineutrino spectra or in the experimental
results. 

We then revised the constraints on the $\theta_{13}$ mixing angle obtained
with the three active neutrino oscillation framework by CHOOZ~\cite{Chooz} 
and KamLAND~\cite{Kamland}. We show that a $\nuebar$ deficit measured 
at any antineutrino experiment with a baseline greater than 1 km could be 
misinterpreted as a hint for a non-zero value of $\theta_{13}$. 
Accounting for a possible $|\Delta m_{\rm new}^2 |$-driven sterile
neutrino oscillation, we revise the constraint on the third mixing
angle, such that $\sin^2(2\theta_{13})<0.11$. Note that the KamLAND 
best fit of the solar neutrino parameters are unchanged. However, the
combination of KamLAND and CHOOZ leads to a best fit value of 
$\sin^2(2\theta_{13})=0.0_{-0.07}^{+0.07}$, in good agreement with the best
fit value extracted from the combined analysis of solar neutrino
data. This relaxes the tension between reactor and solar data 
recently reviewed in Ref.~\cite{theta13review}.

If the existence of a fourth neutrino turns out to be verified we note
that a~7\% reduction of the total flux of solar neutrinos must be
taken into account when confronting with the experimental results. 
In particular the total
neutrino $\nu$ flux measured by SNO~\cite{sno}, 
at 5.14$\pm$0.21~10$^6$~cm$^{-2}$.s$^{-1}$, is now 
in better agreement with the prediction of the (reduced) high-Z Sun
Standard Model model, at 5.25$\pm$0.9~10$^6$~cm$^{-2}$.s$^{-1}$, 
and disfavors the (reduced) low-Z one (4.09$\pm$0.9 10$^6$
cm$^{-2}$.s$^{-1}$). 
We note that the high-Z model is also in good agreement with the data 
from helioseismology, contrarily to the low-Z model~\cite{solaropacity}.

Assuming a hypothetical new neutrino $\nu_{\rm new}$ heavier than
the three active neutrinos we can briefly quantify its contribution to
the effective masses searched for in $\beta$-decay and 
neutrino-less $\beta \beta$-decay experiments, as performed 
in Ref.~\cite{GiuntiReview}. 
The Tritium $\beta$-decay experiments have reported 
$m_{\beta}\lesssim2$~eV (95\%)~\cite{tritium}. 
A fourth neutrino with a mass in the eV range should contribute
more than the active neutrinos, which are expected to have sub-eV masses,
to the signal searched for in direct detection beta decay experiments.
We find that the contribution of the fourth neutrino state fitting the
data analyzed in this article would contribute to $m_{\beta}$
for more than 0.2~eV (95\%~C.L.). This is within the sensitivity range
of the forthcoming KATRIN experiment~\cite{katrin}. 
Assuming Majorana neutrinos, the contribution of the fourth state
would be such that $m_{\beta \beta} \gtrsim0.02$~eV (95\%~C.L.), 
which is above the contribution of the
three active neutrinos and thus disfavors possible cancellations with
the three other active flavors~\cite{betabetabounds,GiuntiReview}.
This contribution is of interest for the forthcoming experiments~\cite{betabetaexp}.

Furthermore we would like to stress that the existence of a fourth
neutrino is slightly favored by some recent cosmological data. 
The effective number of neutrino species fitted by WMAP
and BAO observations~\cite{wmapbao} is N$_{\rm eff}$=4.34$\pm$0.87. 
An analysis of WMAP combined with the Atacama Cosmology Telescope data
~\cite{wmapata} leads to N$_{\rm eff}$=5.3$\pm$1.3,
and an independent analysis of the WMAP 7~years 
data~\cite{wmap7reana} provides N$_{\rm eff}$=4$\pm$1.
Finally a recent analysis based on non-standard big bang nucleosynthesis
~\cite{bbn} leads to N$_{\rm eff}$=3.78$\pm$0.75.
But the compatibility of the sterile neutrino hypothesis exposed in
this paper ($|\Delta m_{\rm new}^2|>$ 1.5 eV$^2$,
\mbox{$\sin^2(2\theta_{\rm new})=0.14\pm0.08$} at 95\%~C.L.) 
with cosmological models and data should be assessed, especially
its contribution to the non-baryonic dark matter of the
Universe, which may be non negligible.

Finally, a clear experimental proof of the presence of this fourth non-standard
neutrino becomes mandatory. This can be given by the imprint on the
energy spectrum in a very short baseline reactor neutrino experiment
or by a new neutrino source experiment in a detector with energy and
spatial resolution.

\section{Acknowledgments}
We would like to thank M.~Fallot, J.~Gaffiot, L.~Giot, J.~Martino,
A.~Porta, V.~Rulhmann-Kleider, \mbox{J.-L.~Sida}, and F.~Yermia for their
careful reading of this manuscript. We are grateful to H.~de Kerret,
J.~Rich, B.~Mansouli\'e and S.~Schoenert for discussions and suggestions.
We thank J.~Beacom and P.~Vogel for fruitful exchange concerning the
precise evaluation of the V-A~inverse beta decay cross section. 

\appendix
\section{Error propagation in reactor experiment analyses}
\label{errprop}
One of the main difficulties in these calculations is the propagation
of the errors on the ILL data from~\cite{SchreckU5_85,SchreckU5Pu9,SchreckPu9Pu1} 
to the final result, i.e. the cross-section per fission for each of the isotopes.
From our previous work in~\cite{LM} we propagated the error on the ILL data
and the conversion procedure to the resulting neutrino spectra. 
However the binning and the energy range
are then those of the ILL data (250 keV bins from 2 to 8 MeV for $^{235}$U).

These data are then fitted to an exponential-polynomial model, see for example
\cite{Huber:2004xh}. This allows to use arbitrarily fine binning, which is
necessary for the correct convolution with the inverse beta cross-section
and also the oscillation probability.
In~\cite{Huber:2004xh} the authors outline a method to propagate the errors
and correlations on the polynomial coefficients to the physical observable.
However as they point out the correlation matrices obtained from the fit
are rather unstable, most coefficients being strongly correlated or anti-correlated
to each other. We found that this made error propagation very difficult.

As an exercise, we performed the following simple Monte-Carlo simulation: using the original
correlation matrix on the converted ILL spectra, we simulated a series of ILL converted spectra,
and fitted each of them using the exponential-polynomial model from~\cite{Huber:2004xh}.
We then evaluated the fitted polynomials in each energy bin, yielding
fitted Monte-Carlo neutrino spectra. 
We then computed the bin-to-bin correlation matrix of these Monte-Carlo spectra.

\begin{figure}[!h]
\begin{center}
\includegraphics[scale=0.4]{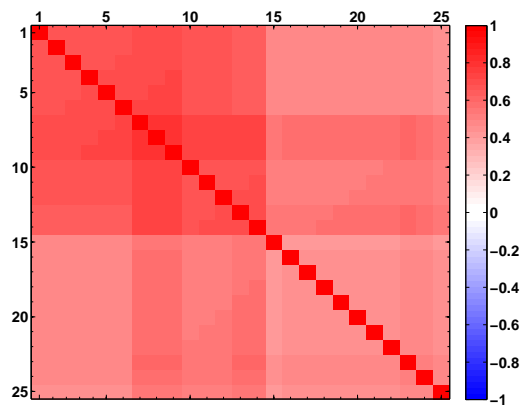}
\caption{\label{f:convcorr} 
Bin-to-bin correlation matrix ($25\times25$~bins, 
2-8~MeV) of converted neutrino spectrum from $^{235}$U,
 including the ILL experimental errors and 
conversion effects from~\cite{LM}.}
\end{center}
\end{figure}

\begin{figure}[!h]
\begin{center}
\includegraphics[scale=0.4]{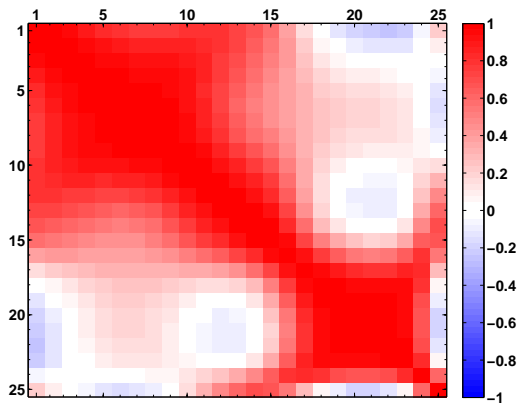}
\caption{\label{f:toycorr} Bin-to-bin correlation matrix ($25\times25$~bins,
  2-8~MeV) of fitted neutrino spectrum from $^{235}$U,
following the toy Monte-Carlo procedure outlined in the text. While the structure
of Figure \ref{f:convcorr} is still present it has been significantly washed out and 
anti-correlation areas have appeared.}
\end{center}
\end{figure}

An example of the result is shown in Figure~\ref{f:toycorr}, along with the
original correlation matrix in Figure~\ref{f:convcorr}. Clearly the polynomial fit induces long-range
correlations and anti-correlations between spectrum bins, and washes out the original correlations.
We concluded that while we could rely on the polynomial fit for the
mean values of the spectra (since the fit residuals are acceptably small), 
we could not use the resulting correlation matrices for error propagation.

\begin{figure}[!h]
\begin{center}
\includegraphics[scale=0.4]{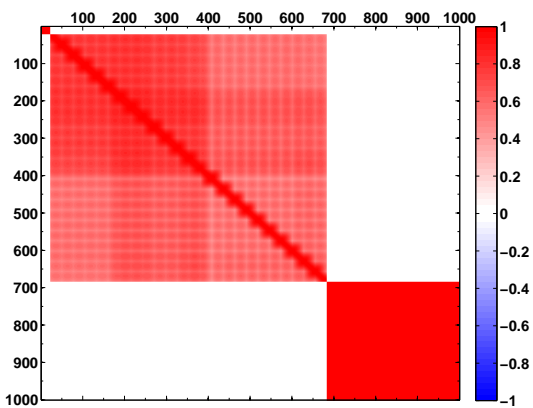}
\caption{\label{f:finalcorr} 
Bin-to-bin correlation matrix ($1000\times1000$~bins, 1.75-10~MeV) 
of the fitted neutrino spectrum from $^{235}$U,
obtained after linear interpolation of the original matrix, 
and extrapolating to a fully correlated
20\% error outside the original binning.
Clearly the experimental correlations from the ILL
measurements are kept with this technique.}
\end{center}
\end{figure}

We therefore choose to propagate the original correlation matrix on our final
binning by interpolating linearly between each bin of the original covariance matrix,
and then computing the resulting correlation matrix.
For bins which were not present in the original energy range, we take a constant
error of 20\%, fully correlated bin-to-bin in order to obtain a conservative upper bound on the error.
This allows us to use any binning over any energy range. Of course the resulting
matrix still has rank~25 over the 2-8~MeV range, 
but since we do not need its inverse this does not limit
our ability to use it for error propagation.
An example of the resulting matrix can be found in Figure~\ref{f:finalcorr} for
$^{235}$U.

Finally, as a cross-check of this method, we computed the error on the cross-section
per fission for each isotope, using our new error propagation method with interpolated
ILL and conversion correlation matrices (column~2 of
table~\ref{tab:crosssec}, reproduced in table~\ref{t:errors} below),
and directly folding the converted ILL spectra with the inverse beta decay cross-section.
Clearly, the results are extremely close. The effect of the error on the extrapolated
part of the spectrum is found to be negligible. We also verified that the resulting error
bars do not significantly depend on the chosen binning.
These calculations validate the use
of the exponential-polynomial fit along with the interpolation technique for error 
propagation on the final spectrum, and allow us to use the binning of our choice 
without affecting the errors.

\begin{table}[t]
\begin{center}
\medskip
\begin{tabular}{c|c|c}
\hline\hline
                   & This work & ILL spectra \\
 Isotope           & 1000 bins, 1.8-10 MeV   & 25 bins, 2-8 MeV \\
\hline
$\sigma^{\rm pred}_{f,^{235}{\rm U}}$  &   6.61$\pm$2.11\%  & 6.61$\pm$2.13\% \\
$\sigma^{\rm pred}_{f,^{239}{\rm Pu}}$ &   4.34$\pm$2.45\%  & 4.33$\pm$2.46\% \\
$\sigma^{\rm pred}_{f,^{241}{\rm Pu}}$ &   5.97$\pm$2.15\%  & 6.02$\pm$2.16\% \\
\hline\hline
\end{tabular}
\caption{\label{t:errors} 
Comparison between cross-sections per fission obtained from
fitted spectra (central column), and from a direct convolution of the
converted ILL spectra (right column). The error
bars are `exact' in the case of the direct convolution, while they are approximate
for the fitted spectrum, obtained from interpolation of the original matrices.}
\end{center}
\end{table}

For $^{238}$U, since the spectrum is obtained from {\it ab initio} calculations,
each bin is given an error ranging from~10\% at low energy to~20\% at high energy.
However, if these errors are uncorrelated, the uncertainty on the overall neutrino rate and hence
the cross-section per fission for $^{238}$U is artificially low.
To avoid this, we consider that there is a fully correlated~10\% error on 
bins from~2 to 5.5~MeV, another fully correlated 15\% error on the block
from~5.5 to 6.75~MeV and another~20\% fully correlated error on the last
bins of the spectrum. 
With this conservative prescription our error on the $^{238}$U
cross-section per fission is~8.2\%.

Finally we derive the total error, $\sigma_{S_{\rm tot}}$, on the antineutrino rate, for a given
core composition. We account for the uncertainties on the cross section
per fission per isotope, and the uncertainties on the averaged fraction of fission per isotopes.
As an example we can consider the case of Bugey-4. The fuel composition
is given in Ref.~\cite{Bugey4}. The new values of the cross sections per
fission per isotope as well as their uncertainties are given in Table 
\ref{tab:crosssec}. The $f_k$ coefficients are taken as correlated, 
following a typical fuel evolution curve, and
such that $\sum_kf_{k}$=1$\pm0.6\%$ to account
for the error on the thermal power of the nuclear core. In a first
case we consider a~10\% relative uncertainty on the fuel composition. 
We obtain a final error on the expected number of events of~2.7\%. 
Reducing the error on the fuel composition to~3.5\% results in a final error of~2.4\%. 

%\appendix
\section{Gallex/Sage, and MiniBooNE reanalysis}
\label{reana}

In this appendix we briefly provide details on our reanalysis of published
data from Gallex  $^{51}$Cr source data
\cite{gallexcr,gallexreana}, SAGE  $^{51}$Cr~\cite{sagecr} 
and $^{37}$Ar source data~\cite{sagear}, and MiniBooNE neutrino data
\cite{miniboone}.

\begin{figure*}[!ht]
\begin{center}
\includegraphics[scale=0.39]{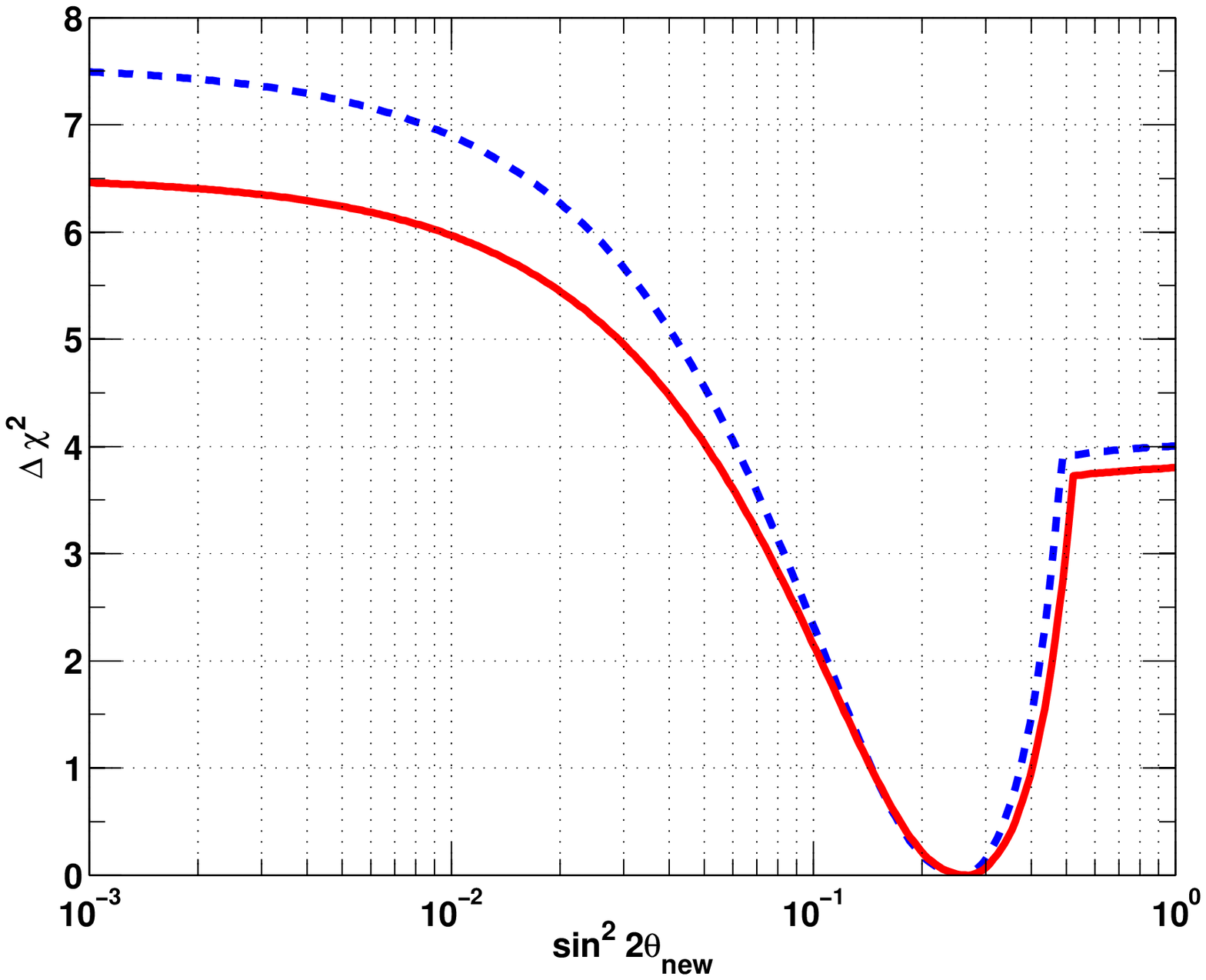}
\includegraphics[scale=0.37]{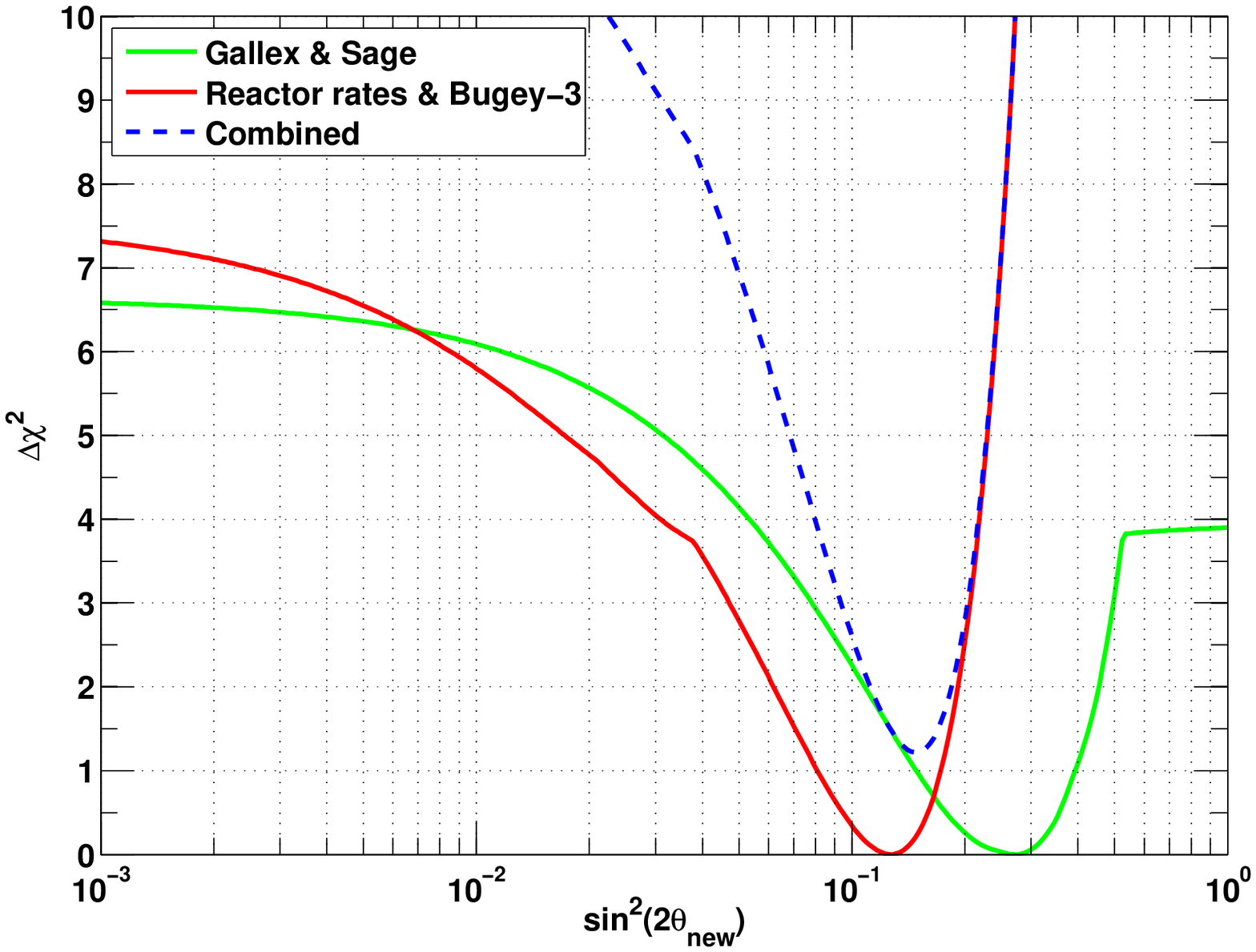}
\caption{\label{f:gallval} (left)
Reanalysis of the gallium calibration data of Gallex and Sage.
$\Delta\chi^2$ projections as a function of~$\sin^2(2\theta)$, without
correlations (dashed) and with correlations as explained in the text.
The dashed contour is very close to that of~\cite{GiuntiReview},
validating this approach. (right) $\Delta\chi^2$ valleys of the
$\sin^2(2\theta_{\rm new})$ parameter of the reactor antineutrino
experimental results compared with the gallium results. Using
the prescription of Ref.~\cite{Pgof} the parameter goodness-of-fit is~27.0\%.
}
\end{center}
\end{figure*}

The Gallex collaboration performed two measurements with two $^{51}$Cr sources
in~\cite{gallexcr}, obtaining the ratio of the measured and expected event rates.
These data were recently reanalyzed~\cite{gallexreana}, providing updated values
of the errors and the ratios.
The SAGE collaboration performed a similar measurement with a $^{51}$Cr source,
and more recently with a $^{37}$Ar source~\cite{sagecr,sagear}.
An analysis of these data in terms of neutrino oscillation was performed
in~\cite{GiuntiReview,GiuntiGallium}.
The data values are $0.95 \pm 0.11$, $0.81 \pm 0.11$, $0.95 \pm 0.12$ and $0.79 \pm 0.09$.
Performing a neutrino oscillation search with these values yields contours
very close to those of~\cite{GiuntiReview}.
However we decided to include possible correlations between these four measurements
in this present work, that were not previously taken into account.
As the two Gallex measurements used the same experimental technique,
we decided to fully correlate their systematic errors, which we understand to
be~5.6\% and~7.4\% respectively. The statistical errors remain of course uncorrelated.
For SAGE we followed a similar procedure, with systematics of~5.7\% and~7.0\%
according to~\cite{sagecr,sagear}.

Our fractional covariance matrix is
\begin{equation}
\left(
\begin{array}{cccc}
  1.31&    0.41 &        0   &   0     \\
  0.41&    1.55 &        0   &   0     \\
  0     &    0      &   1.53 &   0.40\\
  0     &    0      &   0.40 &   1.30\\
\end{array}
\right)
10^{-2}.
\label{covmat2}
\end{equation}

\begin{figure}[!h]
\begin{center}
\includegraphics[scale=0.39]{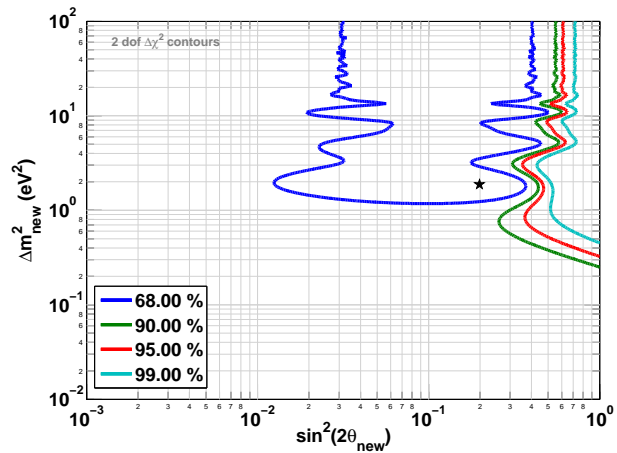}
\caption{\label{f:minib}
Analysis of the MiniBOONE data following a procedure very similar to
that of~\cite{GiuntiReview}. Our results are in good agreement.}
\end{center}
\end{figure}

As expected we obtained slightly more conservative contours than those
of~\cite{GiuntiReview} with this method.

It turns out that the $\sin^2(2\theta_{\rm new})$ best fits obtained from
reactor antineutrino rate results only ($\sin^2(2\theta_{\rm new,best
  fit})\sim 0.10$)
and the gallium results only ($\sin^2(2\theta_{\rm new,best fit})\sim 0.27$)
are slightly different, though with large uncertainties. We quantified
this effect by using the prescription of Ref.~\cite{Pgof}. The
corresponding $\Delta\chi^2$ valleys are displayed on the right panel
of Figure~\ref{f:gallval}. After marginalizing over $|\Delta m_{\rm new}^2|$
(since fitted at the same value in both data sets) the
parameter goodness-of-fit between the two data sets on $\sin^2(2\theta_{\rm new})$
is~27.0\%, indicating no significant tension between neutrino and antineutrino
anomalies. 

We also analyzed the MiniBooNE neutrino data from~\cite{miniboone} following
the method outlined in~\cite{GiuntiReview}. The idea is to fit both electron-like
and muon-like data at the same time.
However, instead of searching for $\nu_{\mu}\rightarrow\nu_e$ appearance,
we search for $\nu_e$ disappearance, while allowing the global normalization
of all the samples (e-like and mu-like) to fluctuate. 

The excess of e-like data in the low energy part of the spectrum is well-fitted
by the combined increase in overall normalization and by the disappearance of
$\nu_e$ induced e-like events.
We used a simplified version of the analysis in reference~\cite{GiuntiReview},
in that we used a covariance matrix independent from the normalization fitting
parameter $f$. We obtain very similar results, although our best fit point
is in a slightly different location, at $\sin^2(2\theta)=0.23$, $\Delta m^2=1.88$~eV$^2$
and $f=1.083$, as can be seen in Figure~\ref{f:minib}.

We have only included this analysis of MiniBooNE's data for completeness.
However in our final statistical significance it has very little impact.

%%%%%%%%%%%%%%%%%%%%%%%%%

\end{document}